\documentclass[useAMS,fleqn,usenatbib]{mn2e}
\usepackage{times}
\usepackage{amsmath}
\usepackage[dvipsnames]{xcolor}
\usepackage{url}
\usepackage{graphicx}
\usepackage{hyperref}
\usepackage{float}
\usepackage{xfrac}

\usepackage{rotating}
\usepackage{epstopdf}
\usepackage{amsmath}
\usepackage{amssymb}
\usepackage{multirow}

\usepackage{breakurl}

\mathchardef\mhyphen="2D

\usepackage{float}

\newcommand{\mbi}[1]{\mbox{\boldmath$#1$}}

\newcommand{\mat}[1]{\mbox{\rm\bf #1}}
\newcommand{\lsim}{\mbox{${\,\hbox{\hbox{$ < $}\kern -0.8em \lower 1.0ex\hbox{$\sim$}}\,}$}}
\newcommand{\gsim}{\mbox{${\,\hbox{\hbox{$ > $}\kern -0.8em \lower 1.0ex\hbox{$\sim$}}\,}$}}

\newcommand{\dd}{{\rm d}}

\def\beqn{\vspace{2mm}
\begin{eqnarray}} 
\def\eeqn{\vspace{2mm} 
\end{eqnarray}}

 \def\la{\langle}

\newcommand{\be}{\begin{equation}}
\newcommand{\ee}{\end{equation}}
\newcommand{\ba}{\begin{eqnarray}}
\newcommand{\ea}{\end{eqnarray}}
\newcommand{\brr}{\begin{array}}
 
\newcommand{\err}{\end{array}}
\newcommand{\bc}{\begin{center}}
\newcommand{\ec}{\end{center}}

\newcommand{\hmpcnosp}{$h^{-1}$\,Mpc}

\chardef\til=`\~

\begin{document}

\title[Inference from biased tracers]{Bayesian inference of cosmic density fields from non-linear, scale-dependent, and stochastic biased tracers}


\author[Metin Ata et al.]
{Metin Ata\thanks{E-mail:mata@aip.de}, Francisco-Shu Kitaura\thanks{E-mail:kitaura@aip.de, Karl-Schwarzschild fellow} \& Volker M\"uller\\
Leibniz Institut f\"ur Astrophysik, 14482 Potsdam, Germany}

\maketitle

\begin{abstract}
We present a Bayesian reconstruction algorithm to generate unbiased samples of the underlying dark matter field from halo catalogues. Our new contribution consists of implementing a non-Poisson likelihood including a deterministic non-linear and scale-dependent bias. In particular we present the Hamiltonian equations of motions for the negative binomial (NB)  probability distribution function. This permits us to efficiently sample the posterior distribution function of density fields given a sample of galaxies using the Hamiltonian Monte Carlo technique implemented in the \textsc{argo} code.
 We have tested our algorithm with the Bolshoi $N$-body simulation at redshift $z=0$, inferring the underlying dark matter density field from subsamples of the halo catalogue  with biases smaller and larger than one. Our method shows that we can draw closely unbiased samples (compatible within 1-$\sigma$) from the posterior distribution up to scales of about $k\sim$1 $h$ Mpc$^{-1}$ in terms of power-spectra and cell-to-cell correlations.  We find that a Poisson likelihood including a scale-dependent non-linear deterministic bias can yield reconstructions with power spectra deviating more than 10\% at $k=0.2$ $h$ Mpc$^{-1}$.
Our reconstruction algorithm is especially suited for emission line galaxy data for which a complex non-linear stochastic biasing treatment beyond Poissonity becomes indispensable.
\end{abstract}

\begin {keywords}
  large-scale structure of Universe -- cosmology: theory -- catalogues -- galaxies: statistics
\end {keywords}

\section{Introduction}

The large-scale structure encodes the key information to understand structure formation and the expansion of the Universe. However, the luminous objects, i.e., the galaxies tracing the large-scale structure, represent only a biased fraction of the total underlying matter governing the laws of gravity.

In the era of precision cosmology, the data analysis methods need to account for the non-linear, biased and discrete nature of the distribution of galaxies to accurately extract any valuable cosmological information. 
This becomes even more important with the advent of the new generation of galaxy surveys. Many of these surveys rely on emission line galaxies, see  WiggleZ\footnote{\burl{http://wigglez.swin.edu.au/site/}} \citep[][]{wigglez2010}, VIPERS\footnote{\burl{http://vipers.inaf.it/}} \citep[][]{guzzo2013}, DESI\footnote{\burl{http://desi.lbl.gov/}}/BigBOSS \citep[][]{bigboss2011}, DES\footnote{\burl{http://www.darkenergysurvey.org}} \citep[][]{des2013}, LSST \footnote{\burl{http://www.lsst.org/lsst/}} \citep[][]{lsst2012}, J-PAS\footnote{\burl{http://j-pas.org/}} \citep[][]{jpas2014}, 4MOST\footnote{\burl{http://www.aip.de/en/research/research-area-ea/research-groups-and-projects/4most}} \citep[][]{4most} or Euclid\footnote{\burl{http://www.euclid-ec.org}} \citep[][]{2009ExA....23...39C,euclid2009}.  
These objects provide denser sampled volumes, deeply tracing the non-linear cosmic web structure. Moreover they introduce a more complex biasing, covering a wider range of galaxy masses, as compared to, e.g., luminous red galaxies.

We present in this work a Bayesian approach designed to deal with non-linear stochastic biased tracers. 
We consider non-Poisson  probability distribution functions (PDFs) for the likelihood modeling the distribution of galaxies. In this way we account for the over-dispersion (larger dispersion than Poisson) of galaxy counts.
 We model the expected galaxy number density relating it to the dark matter density through a non-linear scale-dependent expression extracted from $N$-body simulations (see \cite{Cen-Ostriker-93}; \cite{delaTorre2012}; \cite{2014MNRAS.439L..21K}; \cite{2014MNRAS.441..646N}; \cite{2014arXiv1407.2637A}).  In this way we extend the works based on the Poisson and linear bias models (\cite{kitaura}; \cite{kitaura_log}; \cite{jasche_hamil}; \cite{jasche_2lpt}) following the ideas presented in \cite{kitaura_lapalma}.
In particular, we implement these improvements in the \textsc{argo} Hamiltonian-sampling code, which is able to jointly infer density, peculiar velocity fields and power-spectra  (\cite{kitaura_lyman}).
For the prior distribution describing structure formation of the dark matter field we use the lognormal assumption  (\cite{Coles1991}).  We note however, that this prior can be substituted by another one, e.g., based on Lagrangian perturbation theory (see \cite{Kitaura_kigen}; \cite{jasche_2lpt}; \cite{2013ApJ...772...63W}; \cite{hesscs}).  Alternatively, one can extend the lognormal assumption in an Edgeworth expansion to include higher order correlation functions (\cite{colombi}; \cite{kitaura_skewlog}).
We show in this work that our likelihood model is able of yielding unbiased dark matter field reconstructions on $\sim6\, h^{-1}$ Mpc scales based on $N$-body simulations. 

This paper is structured as follows. In \S~\ref{sec:bayes}, we describe our statistical approach. Then we present our numerical tests (\S~\ref{sec:results}) and finally we present our conclusions (\S~\ref{sec:conc}).

\section{Method}

\label{sec:bayes}

To infer the dark matter density field from biased tracers such as galaxies or halos, one has to define a target distribution, called posterior PDF. 
We use the Bayesian framework to express the posterior distribution based on a model PDF for the data, the likelihood, and a model PDF for the signal, the prior. We then recap the Hamiltonian sampling technique used in this work to sample from such a PDF, and present the Hamiltonian equations of motions for our model.

\subsection{Bayesian approach: the posterior distribution function}

In this work we will restrict ourselves to the reconstruction of dark matter fields given a set of biased tracers in real-space. We note that redshift-space distortions can be corrected within a Gibbs-sampling scheme and leave this additional complication for a later work \citep[see][]{kitaura,kitaura_lyman}.

Let us divide the volume under consideration into a grid with $N_{\rm C}$ cells. Our input data vector is given by the number counts of halos or galaxies per cell $\mbi N_{\rm G}$ and the desired signal is the dark matter density $\mbi\delta_{\rm M}$. In addition, we need to assume some model for the dark matter distribution $\mathcal M(\mbi\delta_{\rm M})$ and for the bias relating the number counts of galaxies to the underlying dark matter distribution $\mathcal B(\mbi N_{\rm G}|\mbi\delta_{\rm M})$, including the nonlinear deterministic and stochastic parameters (see \S \ref{sec:stochastic} and \S \ref{sec:deterministic}). We will assume in this work that there is one population of tracers and that they can be described with a set of bias parameters. Nevertheless, we show in an appendix (see \S \ref{sec:multi}) that we could sub-divide the sample into various tracer types, each one with its own bias parameters, and combine them in a multi-tracer analysis in a straightforward way, within the methodology presented here.
The posterior distribution function of dark matter fields given $\mbi N_{\rm G},\mathcal M(\mbi\delta_{\rm M})$, and $\mathcal B(\mbi N_{\rm G}|\mbi\delta_{\rm M})$ can be expressed within the Bayesian framework as the product between the prior $\pi$ and the likelihood $\mathcal L$ up to a normalization
\ba
\lefteqn{\mathcal P(\mbi\delta_{\rm M}|\mbi N_{\rm G},\mathcal M(\mbi\delta_{\rm M}),\mathcal B(\mbi N_{\rm G}|\mbi\delta_{\rm M}))\propto}\\&&\pi(\mbi\delta_{\rm M}|\mathcal M(\mbi\delta_{\rm M}))\times\mathcal L(\mbi N_{\rm G}|\mathcal B(\mbi N_{\rm G}|\mbi\delta_{\rm M}))\nonumber\,.
\ea

\subsection{The Likelihood: stochastic bias}
\label{sec:stochastic}

Let us start defining the likelihood, which models the statistical nature of the data.
First we assume that the biasing relation is known and define the expected number count by $\mbi\lambda\equiv\langle\mbi N_{\rm G}\rangle_{\rm G}$,  where $\langle \{\dots\} \rangle_{\rm G}\equiv \sum_{N_{\rm G}=0}^\infty \mathcal L(\mbi N_{\rm G}|\mathcal B(\mbi N_{\rm G}|\mbi \delta_{\rm M}))\{\dots\}$ denotes the ensemble average over the halo or galaxy realization. 
At this point we need to introduce the stochastic  biasing parameters $\{p_{\rm SB}\}$, necessary to model the deviation from the Poisson distribution $\mathcal L(\mbi N_{\rm G}|\mathcal B(\mbi N_{\rm G}|\mbi\delta_{\rm M}))=\mathcal L(\mbi N_{\rm G}|\mbi\lambda,\{p_{\rm SB}\})$. 
The  positive (negative) correlation of halos at sub-grid resolution introduces over- (under-) dispersed distributions depending on the halo population and density regime \citep[see][]{2001MNRAS.320..289S,2002MNRAS.333..730C,2014MNRAS.441..646N}. This effect was already predicted by \citet{peebles-80}.
We focus on modeling over-dispersion, as under-dispersion is a sub-dominant effect, only present for very massive objects \citep[see][]{2012PhRvD..86h3540B,baldauf2013}.
We note, that stochastic bias has  been studied in a number of works, more or less explicitly  \citep[see, e.g., ][]{1974ApJ...187..425P,1985MNRAS.217..805P,bbks1986,Fry1993,1996MNRAS.282..347M,1999ApJ...520...24D,1999MNRAS.304..767S,2000MNRAS.318..203S,2002ApJ...575..587B,2007PhRvD..75f3512S,2010PhRvD..82j3529D,2011PhRvD..83j3509B,2011A&A...527A..87V,2012MNRAS.421.3472E,2012PhRvD..85h3509C,2012PhRvD..86h3540B,baldauf2013}.

For a given distribution function $f(\lambda,N,\{p_{\rm SB}\})$ with expectation value $\lambda$, observed number count $\mbi N_{\rm G}$ and the set of stochastic bias parameters $\{p_{\rm SB}\}$, the likelihood can be written as follows
\begin{equation}
\mathcal{L}(\mbi N_\mathrm{G}|\mbi \lambda,\{p_{\rm SB}\})= \prod_{i=1}^{N_{\rm C}} f(\lambda_i,N_i,\{p_{\rm SB}\}) \, .
\end{equation}
The product is computed over the number of cells $N_{\rm C}$, that corresponds to the number of dimensions of the problem.

Let us consider the negative binomial distribution  (NB)   and the gravitational thermodynamics distribution  by \cite{1984ApJ...276...13S}  (GT), to describe the deviation from Poissonity, which require a single stochastic bias parameter $\beta$ and $b$, respectively.
  The Poisson  $f_{\rm P}$ the NB $f_{\rm NB}$ and the GT $f_{\rm GT}$ distributions are written as
\begin{eqnarray}
f_{\rm P}(\lambda,N) &=&\frac{ \mathrm{e}^{-\lambda} \lambda^N}{N!} \, , \\
 f_{\rm NB}(\lambda,N,\beta) &=& \frac{\lambda^{N}}{N!} \frac{\Gamma(\beta+N)}{\Gamma(\beta)(\beta+\lambda)^{N}}\frac{1}{\left(1+\frac{\lambda}{\beta}\right)^\beta} \, , \\
 f_{\rm GT}(\lambda,N,b) &=& \frac{\lambda}{N!}{\rm e}^{-\lambda(1-b)-bN}(1-b)\, [\lambda(1-b)+bN]^{N-1} \nonumber\, , \\ 
\end{eqnarray}
respectively.
The parameters $\beta$ and $b$ are connected to the expectation value
$\lambda$ and the variance $\sigma^2$ by $\beta =
\lambda^2/(\sigma^2-\lambda)$ and $b=1-\sqrt{\lambda/\sigma^2}$,
respectively. This implies that the over-dispersion term shows a quadratic and
a linear dependence on the expected halo number count $\lambda$ for the NB
$\sigma_{\rm NB}^2=\lambda+\lambda^2/\beta$ and the GT $\sigma_{\rm
  GT}^2=\lambda/(1-b)^2=\lambda+\lambda\,b(2-b)/(1-b)^2$ case,
respectively. To obtain a different dependence, one could take the NB
expression and include a dependence of $\beta$ on $\lambda$. For
$\beta\propto\lambda$ we find that the NB and the GT PDFs are equivalent in
terms of over-dispersion. For this reason we will focus in the following on the NB
PDF and leave a study on the $\lambda$ dependence for a later work. One could investigate these dependencies for different population of halos, by taking for instance, ensembles of high resolution simulations like those in \citet[][]{2012arXiv1210.7871A}.

\subsection{Link between prior and likelihood: deterministic bias}
\label{sec:deterministic}

Let us now define the link between the likelihood and the desired signal, i.e., the dark matter density field.
This  is given by the deterministic bias relating the expected number counts to the dark matter over-density field, which is in general non-linear, scale dependent and non-local. We note that non-locality introduces a scatter which can be absorbed in the stochastic bias \cite[see discussion in][]{kitauraBispetrum}.

One could expand  the dark matter overdensity field $\delta_{\rm
  M}\equiv\rho_{\rm M}/\bar{\rho}_{\rm M}-1$ (with $\bar{\rho}_{\rm M}$ being
the mean dark matter density) in a Taylor series
\citep[][]{Fry1993}. Alternatively one could follow
\citet[][]{Cen-Ostriker-93} and expand the series based on the logarithm of
the density field (avoiding in this way negative densities allowed in the
previous expansion). This model to linear order, corresponding to a
power-law of the dark matter field, has been proposed for resolution augmentation
of $N$-body simulations \citep[see][]{delaTorre2012}. The power law can be interpreted as a linear bias factor in Lagrangian space which undergoes gravitational evolution within the lognormal approximation. It has recently been found that the bias is very well fit by a compact relation including an exponential cut-off:  $\rho_{\rm h}\propto\rho_{\rm M}^\alpha\,\exp\left[-\left(\frac{\rho_{\rm M}}{\rho_\epsilon}\right)^{\epsilon}\right]$ \citep[see][]{2014MNRAS.441..646N}, which can be approximated by a Heaviside step-function \citep[$\theta(\delta_{\rm M}-\delta_{\rm{th}})$, $=0$ if $\delta_{\rm M} < \delta_{\rm{th}}$, else $= 1$, see][]{2014MNRAS.439L..21K}. This model is in concordance with the \citet{1974ApJ...187..425P} and the peak background split formalism \citep[see, e.g.,][]{Kaiser84,bbks1986,Cole89,1996MNRAS.282..347M,Sheth01}, which permit the formation of haloes only above a certain density threshold. It has recently been shown to be a crucial ingredient in the halo three point statistics \citep[][]{kitauraBispetrum}. 
The deterministic bias model including the density power-law and the density cut-off is given by the following expression for the expected halo/galaxy density:
\begin{equation}
{\lambda_i\equiv \langle \rho_{{\rm G}i} \rangle_{\rm G}=}  f_{\bar{N}}w_i (\rho_{{\rm M}i})^\alpha \,\theta(\delta_{\rm M}-\delta_{\rm{th}})  \,,
\label{eq:detbias}
\end{equation}
where $w_i$ is the completeness at each cell $i$.
In this way our method is able of dealing with incomplete data samples as well \citep[see also][]{kitaura_log,jasche_hamil,kitaura_lyman}. 
The normalization ensuring a particular mean number count $\bar N$ is given by
\begin{equation}
f_{\bar{N}} = \frac{\bar{N}}{ \left \langle {  w_i (\rho_{{\rm M}i})^\alpha  \theta(\delta_{\rm Mi}-\delta_{\rm{th}})  } \right \rangle_{\rm M} }\,,
\end{equation}
where $\langle \{\dots\} \rangle_{\rm M}\equiv \int \dd \mbi \delta_{\rm M}\mathcal \pi(\mbi \delta_{\rm M}|\mathcal M(\mbi \delta_{\rm M}))\{\dots\}$ denotes the ensemble average over all possible dark matter realizations (cosmic variance). 

\subsection{The Prior: statistics of dark matter fields}

Let us define now the model describing the distribution of dark matter density fields $\mathcal M(\mbi \delta_{\rm M})$. The Gaussian assumption has since long been commonly used in the literature. This assumption is for instance present in the Wiener reconstruction method  \citep[see, e.g., ][]{zaroubi,kitaura_sdss}.
Nevertheless, it is well known that gravity induces deviations from Gaussianity.
For this reason a non-Gaussian model is essential to exploit the accuracy of our likelihood model. For the sake of simplicity we will restrict this work to the lognormal assumption (see \citet[][]{Coles1991}, and its implementation within a Bayesian context \citet{kitaura_log}). We note, however, that any structure formation model can be implemented at this point as has been shown in a series of recent works \citep[see][]{Kitaura_kigen,jasche_2lpt,2013ApJ...772...63W,hesscs,wang2}. 
Moreover, the lognormal assumption has recently been shown to satisfy
sufficient statistics \cite{2014arXiv1407.0245C}, and thereby extract the
maximal cosmological information. However, it is known that the lognormal
approximation fails at modeling the large-scale structure in the low density
regime \citep[see, e.g.,][]{colombi}. However, it has been shown to be a good
approximation for the moderate to high density regime \citep{kitaura_lapalma}.
We leave an extension with more complex models like those including higher
order statistics for later work \citep[][]{kitaura_skewlog}.

Let us define the signal distribution $\mbi s$ as a logarithmic transformation of the matter density field
\begin{equation}
\mbi s=\log(1+\mbi\delta_{\rm M}) -\mbi \mu \, ,
\end{equation}
 with the mean field $\mbi \mu=\langle\log(1+\mbi \delta_{\rm M})  \rangle_{\rm G}$  ensuring that the mean of $\mbi s$ vanishes, $\langle \mbi s \rangle =0 $. The Gaussian prior for the logarithmically transformed density field is then given as
\begin{equation}
\label{prior}
 \pi(\mathbf{S})= \frac{1}{\sqrt{(2\pi)^{N_{\mathrm{C}}}\mathrm{det}(\mathbf{S})}} \exp \left( -\frac{1}{2} {\mbi s}^{\dagger} \mathbf{S}^{-1} \mbi s\right) \, , 
\end{equation}
where the covariance matrix $\mathbf{S}\equiv\langle \mbi s^\dagger\mbi s\rangle_{s}$ (or power spectrum in Fourier-space) depends on the set of cosmological parameters $\{p_{\rm C} \}$, where $\langle \{\dots\} \rangle_{s}\equiv \int \dd \mbi s\,\mathcal \pi(\mbi s|\mat S)\{\dots\}$ denotes the ensemble average over all possible lognormal fields. In practice, we assume a linear power spectrum and the corresponding covariance matrix.

\subsection{Sampling from the posterior distribution function}
\label{sec:sampling}






%

\begin{figure*}
\parbox{5.5cm}{
\hspace{-3.cm}
\includegraphics[width=5.5cm]{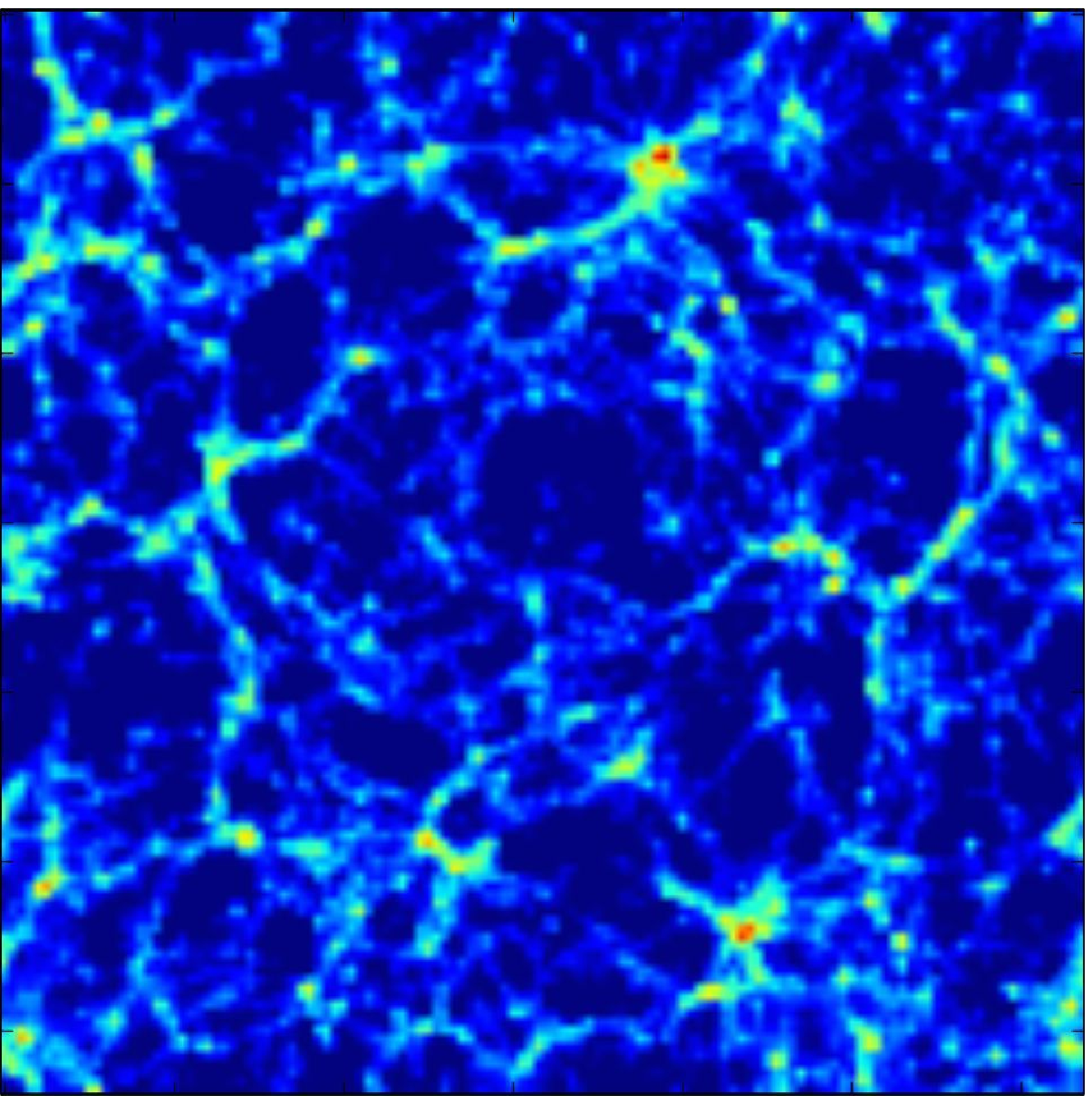}
}
\parbox{5.5cm}{

\begin{tabular}{ccc}

\hspace{-3.cm}

\hspace{-.0cm}
\includegraphics[width=5.5cm]{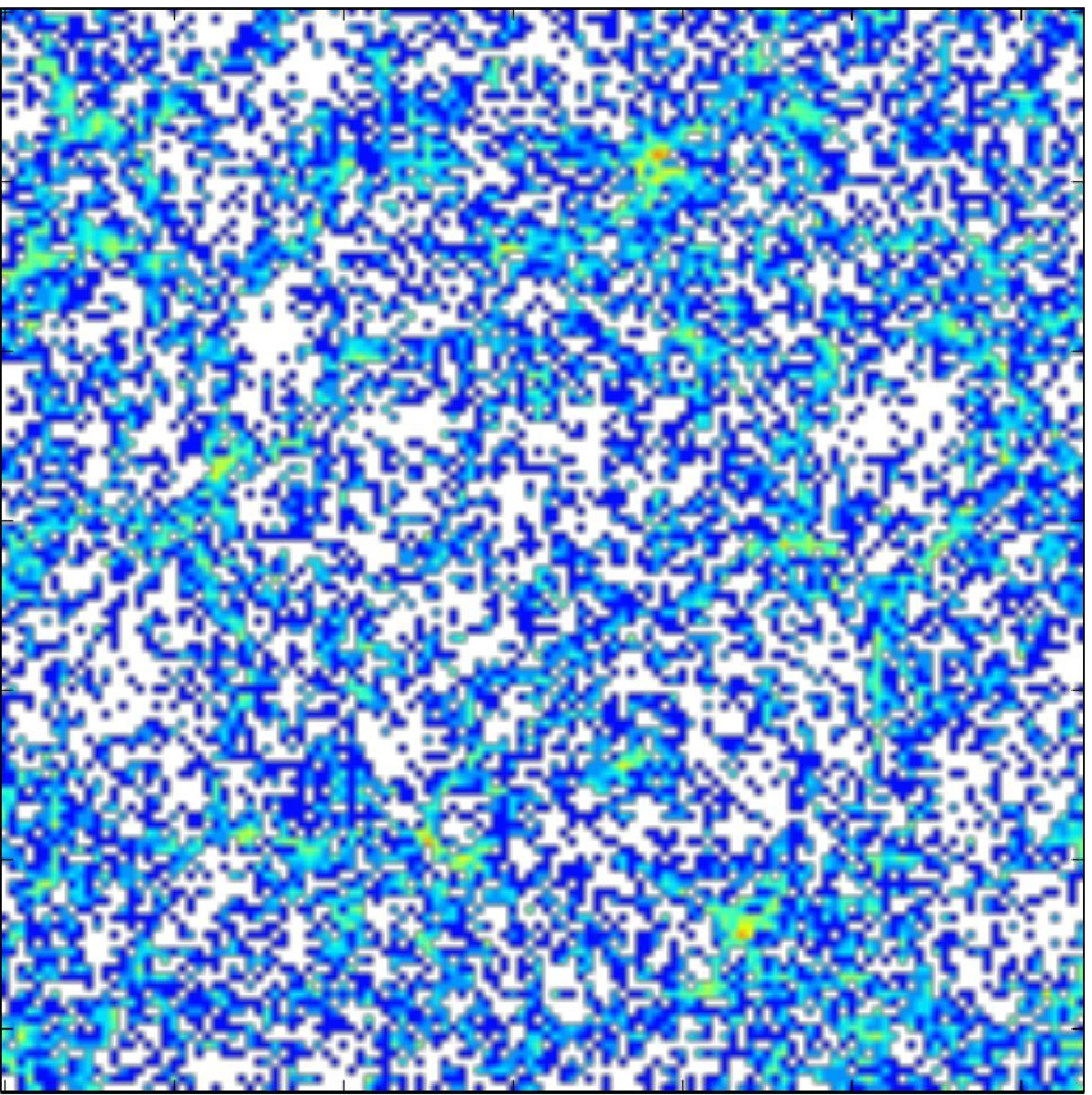}

\hspace{-.0cm}
\includegraphics[width=5.5cm]{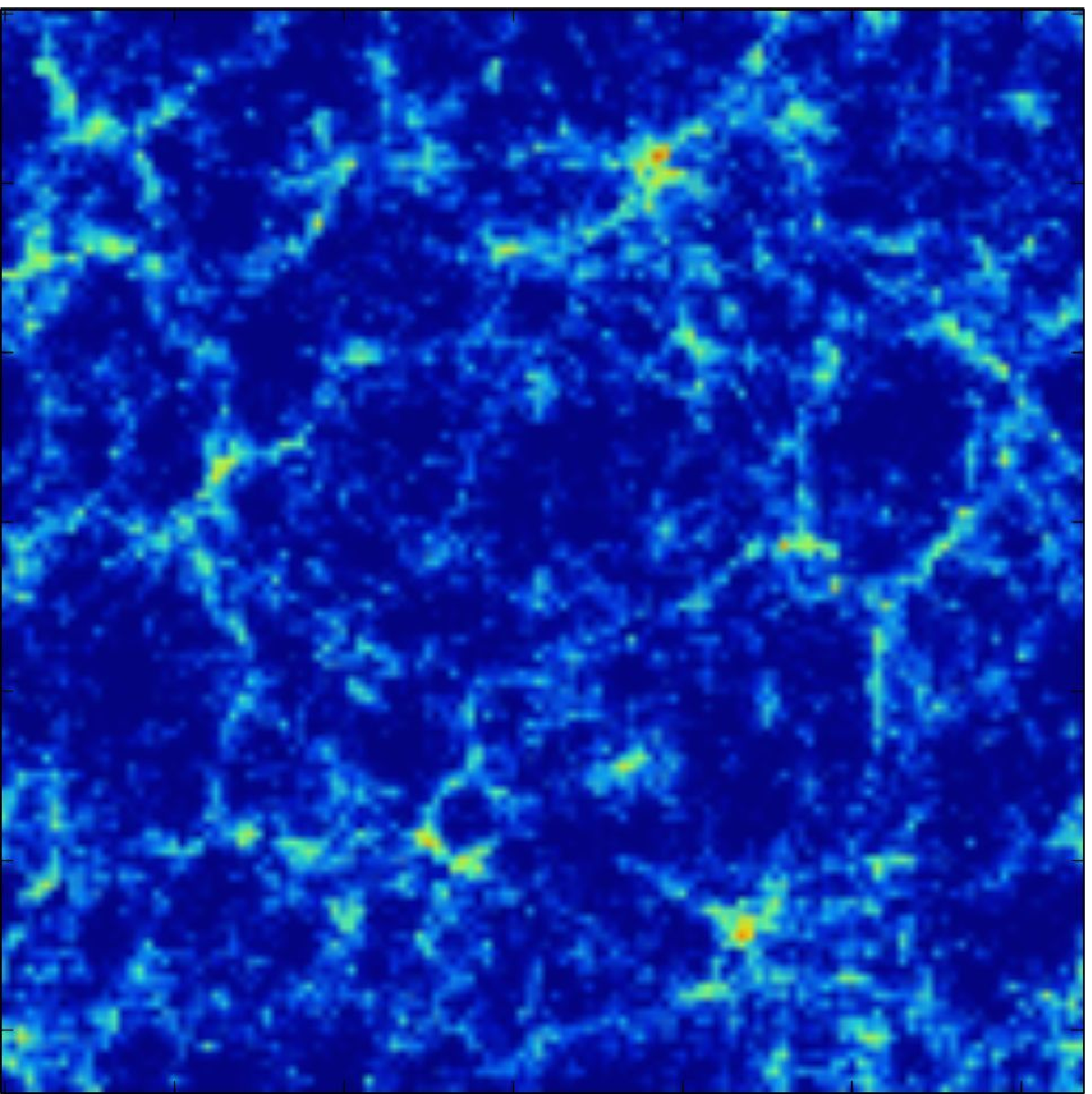}

\includegraphics[width=.40cm]{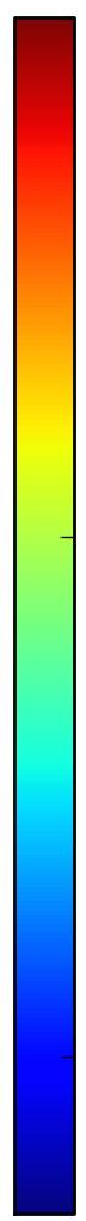}

\put(0,153){\text{$10^3$}}
\put(0,87.5){\text{$10^2$}}
\put(0,19.5){\text{10}}
\\
\hspace{-3.cm}

\hspace{-.0cm}
\includegraphics[width=5.5cm]{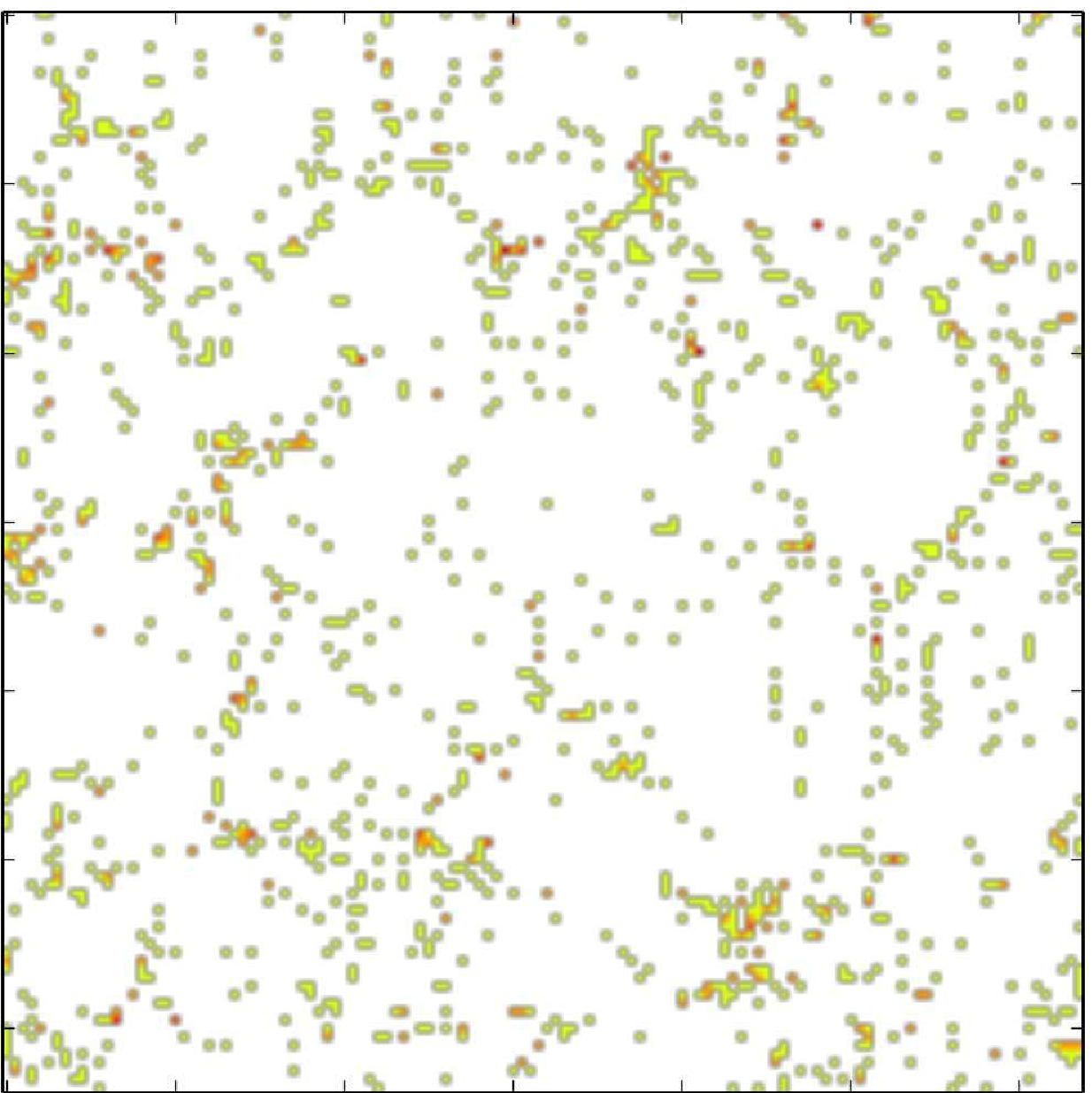}

\hspace{-.0cm}
\includegraphics[width=5.5cm]{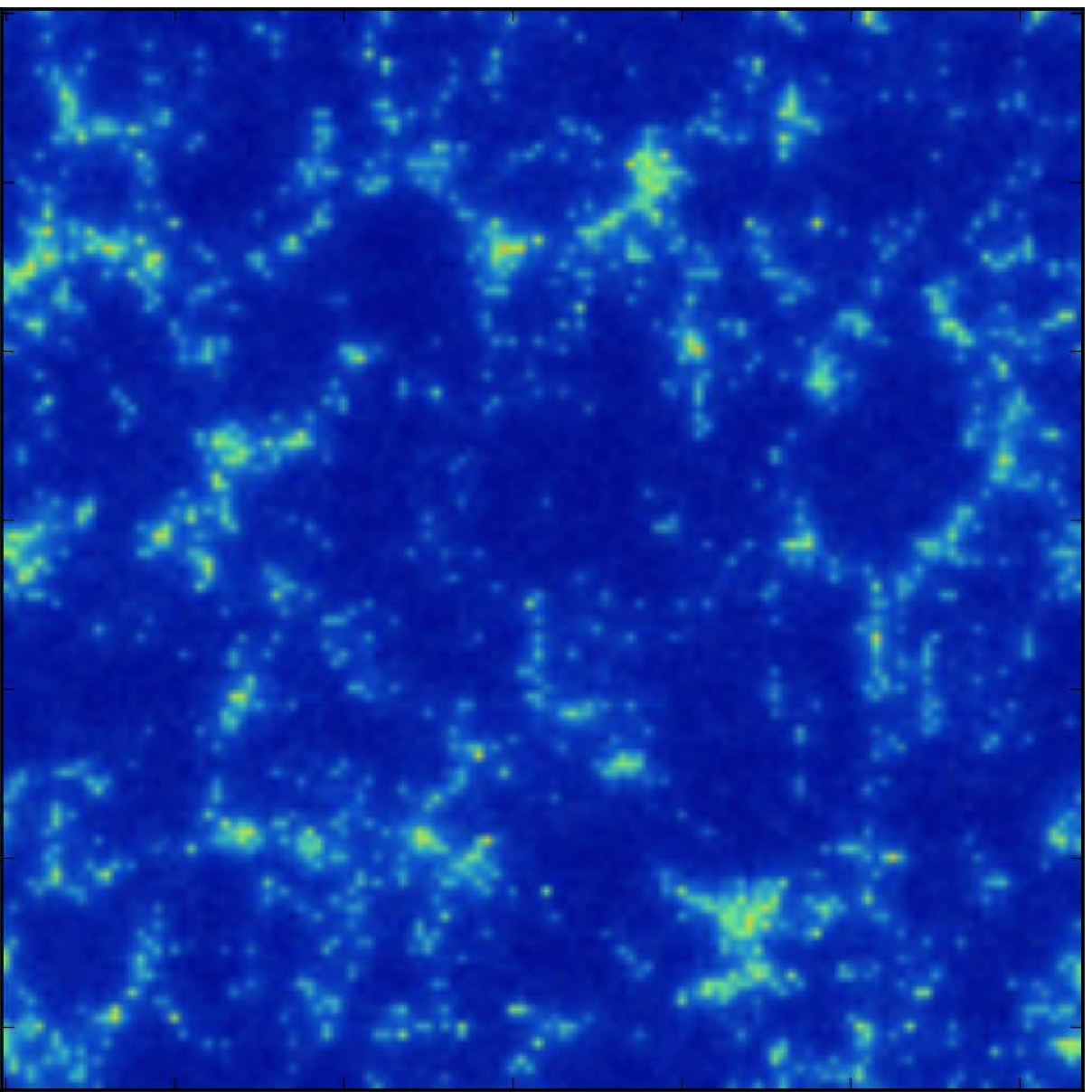}

\includegraphics[width=.40cm]{cb_sliceb}

\put(0,153){\text{$10^3$}}
\put(0,87.5){\text{$10^2$}}
\put(0,19.5){\text{10}}

\end{tabular}
} 
\vspace{0.5cm}
\caption{\label{fig:slice} Slices of the density field ($1+\delta$) with
  thickness $\sim10$ {\hmpcnosp}  from a volume of 250 {\hmpcnosp} side for
  {\it left panel}:   the dark matter field  from the Bolshoi simulation
  (about $9\times 10^9$ dark matter tracers), {\it middle panel, top}: halo
  catalogue $S_1$ from the Bolshoi simulation with $2\times 10^5$ matter
  tracers,  {\it middle panel, bottom}: halo catalogue $S_2$ from the Bolshoi simulation created from haloes of at least $3\times10^{12}\, \rm M_{\odot}$, resulting in $3\times10^4$ matter
  tracers, {\it right panel, top}:  the \textsc{argo} NB reconstruction of $S_1$, and {\it right panel, top}: the \textsc{argo} NB reconstruction of $S_2$. The \textsc{argo} reconstructions have been averaged over 10000 Hamiltonian iterations. The colour code indicates the density  $1+\delta$.}
\end{figure*}

To sample from the posterior distribution we rely on the Hamiltonian Monte Carlo sampling technique (HMC), which is a hybrid  Markov Chain Monte Carlo (MCMC) approach that uses physically motivated equations of motion to explore the phase space avoiding inefficient random walks.
Let us recap here the basics of Hamiltonian sampling. For more details we refer to \cite{duane}, \cite{neal1993} and more recently \cite{2012arXiv1206.1901N}). For applications to astronomy we refer to, e.g., \cite{2008MNRAS.389.1284T}, \cite{jasche_sdss}, and \cite{kit2mrs}.
Here we briefly point to the main features and also emphasize the modifications we implemented. 
The HMC approach treats the problem as a thermodynamical system. The contact with a heat bath moves the system into equilibrium, i.e., to the statistical space in which samples are drawn from the posterior distribution $\mathcal{P}(\mbi{x})$. The key idea of the HMC approach consists of moving the system by solving the Hamiltonian equations of motions, which involve a stochastic kinetic term.
Let us define in this physical analogy the potential ($U$), kinetic ($K$) and Hamiltonian  ($\mathcal{H}$) energy through the following relations
\begin{eqnarray}
\label{ham1}
 U(\mbi{x}) &=& -\ln{\mathcal{P}(\mbi{x})}\, , \\
\label{ham2}
\mathcal{H}(\mbi{x},\mbi{p})&=&U(\mbi{x})+K(\mbi{p})\, ,
\end{eqnarray}
where $U(\mbi{x})$ is the potential function at the coordinate vector $\mbi{x}$ and $K(\mbi{p})$ is the kinetic term of the momentum $\mbi{p}$ of the form ${\sum_{i,j} \frac{1}{2} p_i M^{-1}_{ij} p_j}$. 
Combining Eqs. \ref{ham1} \& \ref{ham2}, we see that the target distribution $\mathcal{P}(\mbi{x})$ can be inferred from
\begin{equation}
\label{exp}
\exp({-\mathcal H}) = \mathcal{P}(\mbi{x})  \cdot \exp\left({-\sum_{i,j} \frac{1}{2} p_i M^{-1}_{ij} p_j}\right).
\end{equation}
This Eq. shows that drawing samples from $\exp(-\mathcal H)$ yields the desired distribution if one is withdrawing the kinetic term, by marginalizing over the momenta. The momenta  are drawn from a multivariate Gaussian distribution with covariance matrix $M_{ij}$.
In order to explore the high dimensional phase space we need to evolve the system. Therefore we solve the Hamiltonian equations of motion
\begin{eqnarray}
\label{eom1}
\frac{d x_i}{dt}&=&\frac{\partial \mathcal H}{\partial p_i}\\
\label{eom2}
\frac{d p_i}{dt}&=&-\frac{\partial \mathcal H}{\partial x_i} = -\frac{\partial U}{\partial x_i} \, .
\end{eqnarray} 
Having fully evolved the initial position $(\mbi{x_0,p_0})$ of the system, we apply a criterion to accept or withdraw the new point in phase space $(\mbi{x_1,p_1})$ that writes
\begin{equation}
\label{accept}
P_{\mathrm{Accept}}={\rm min} \left[1,\exp\left(-\mathcal
  H(\mbi{x_1},\mbi{p_1})+\mathcal H(\mbi{x_0},\mbi{p_0})\right)  \right]\,.
\end{equation}

\subsection{Hamiltonian Equations of motion}
\label{sec:ham}

As shown in Eq. \ref{ham1}~,\ref{eom1}~,\ref{eom2} and \ref{accept}, we need to calculate the logarithm of the target distribution $\mathcal{P}(\mbi{x})$ and its gradient term. Let us summarize below the  potential energy terms and the corresponding analytic gradient expressions required in our work.

\subsubsection{Gaussian prior}

We calculate the negative logarithm of Eq. \ref{prior} as
\begin{eqnarray}
-\ln \pi &=& -\ln \left[ \frac{1}{\sqrt{(2\pi)^{N_{\mathrm{C}}}\mathrm{det}(\mathbf{S})}} \exp \left( -\frac{1}{2} {\mbi s}^{\dagger} \mathbf{S}^{-1} \mbi s\right)  \right] \nonumber \\
-\ln \pi &=&  -\frac{1}{2} {\mbi s}^{\dagger} \mathbf{S}^{-1} \mbi s - c \, ,
\end{eqnarray}
where $c$ incorporates all constant terms of the normalization. \\
The derivative w.r.t. the signal ${\mbi s}$ writes
\begin{eqnarray}
 - \frac{\partial}{\partial {\mbi s}} \ln \pi &=& - \frac{\partial}{\partial {\mbi s}}  \ln \left[ \frac{1}{\sqrt{(2\pi)^{N_{\mathrm{C}}}\mathrm{det}(\mathbf{S})}} \exp \left( -\frac{1}{2} {\mbi s}^{\dagger} \mathbf{S}^{-1} \mbi s\right)  \right] \nonumber  \\
  - \frac{\partial}{\partial {\mbi s}} \ln \pi &=&   \mathbf{S}^{-1} \mbi s
\end{eqnarray} 
Here we substituted the dummy variable $\mathbf{x}$ with the logarithmic signal variable ${\mbi s}$ we want to evaluate.
We note, that more complex schemes connecting the initial conditions with the final gravitationally evolved density fields assume this prior for the primordial fluctuations, but include the  Lagrangian to Eulerian mapping in the likelihood (see \cite{Kitaura_kigen}; \cite{jasche_2lpt}; \cite{2013ApJ...772...63W}; \cite{hesscs}).
We will show below the likelihood expressions for the lognormal case, but include more general expressions in the appendix (see \S \ref{app:grads}).

\subsubsection{Poisson Likelihood}

The terms for the Poisson likelihood write as follows
\begin{eqnarray}
\mathcal{L}(\mbi N_\mathrm{G}|\mbi \lambda) &=& \prod_{i}^{N_{\mathrm{C}}} \frac{ \mathrm{e}^{-\lambda_i} {\lambda_i}^{N_i}}{N_i!} \nonumber \, , \\
-\ln \mathcal{L} &=& \sum_i \left(\lambda_i -N_i \ln \lambda_i - c \right) \, .
\label{poissonlike1}
\end{eqnarray}
Finally the derivative w.r.t. to signal variable ${\mbi s}$ writes as 
\be
\label{poissonlike}
- \frac{\partial \mathcal{L}}{\partial s_i}= \alpha \lambda_i \cdot \left[1 - \frac{N_i}{\lambda_i} \right] \, . 
\ee

\subsubsection{Non-Poisson Likelihood: Negative Binomial}

We calculate the corresponding terms for the NB distribution:

\begin{eqnarray}
\mathcal{L}_{\rm NB}(\mbi N_\mathrm{G}|\mbi \lambda,\beta) &=& \prod_{i}^{N_{\mathrm{C}}}   \left[ \frac{{\lambda_i}^{N_i}}{N_i!} \frac{\Gamma(\beta+N_i)}{\Gamma(\beta)(\beta+\lambda_i)^{N_i}}\frac{1}{\left(1+\frac{\lambda_i}{\beta}\right)^\beta} \right]
  \nonumber , \\ 
-\ln \mathcal{L}_{\rm NB} &=& \sum_i \left(-N_i\ln \lambda_i
+N_i\ln(\beta+\lambda_i)\right . \\ && \left . +
\beta\ln(1+{\lambda_i}/{\beta})-c \right) . 
\end{eqnarray}
Corresponding to Eq. \ref{poissonlike} we can write the derivative of these likelihood functions as

\be
- \frac{\partial \mathcal{L}_{\rm NB}}{\partial s_i}=  \alpha  \lambda_i \cdot \left[ \frac{1}{\frac{\lambda_i }{\beta }+1}+\frac{N_i}{\beta +\lambda_i }-\frac{N_i}{\lambda_i } \right ] \, . 
\ee

\subsection{Leap frog scheme}

To explore the phase space and solve iteratively the Hamiltonian equations of motion we use the leapfrog scheme
\begin{eqnarray}
p_i\left(t+\frac{\epsilon}{2}\right) &=& p_i(t)-\frac{\epsilon}{2} \frac{\partial U(\mbi x)}{\partial x_i } \, , \\
x_i(t+\epsilon) &=& x_i(t) -\frac{\epsilon}{m_i} \,p_i\left(t + \frac{\epsilon}{2}\right) \, , \\
p_i\left(t+\epsilon\right) &=& p_i\left(t + \frac{\epsilon}{2}\right)-\frac{\epsilon}{2} \frac{\partial U(\mbi x)}{\partial x_i } \,,
\end{eqnarray}
with $n$ being the number of steps, $\epsilon$ the step size and  the total pseudo time given by $\tau= n\epsilon$.

\begin{figure}
\begin{tabular}{c}
\hspace{0.5cm}
\includegraphics[width=7.cm]{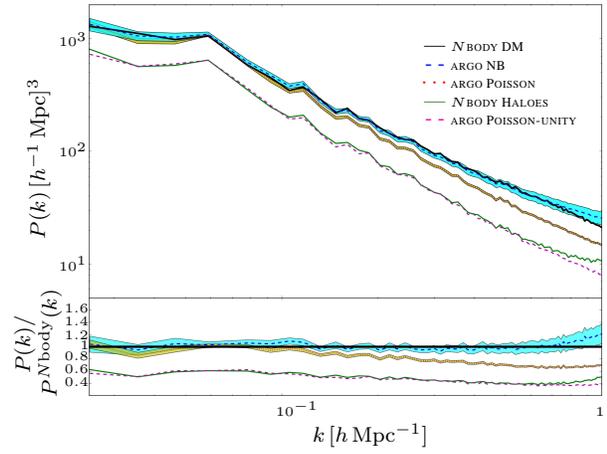}


\put(-70,133){\textcolor{black}{---}}
\put(-60,133){\text{\tiny \textsc{$N$body DM}}}
\put(-69.5,126){\textcolor{blue}{- -}}
\put(-60,126){\text{\tiny \textsc{argo} \textsc{NB}}}
\put(-71,119){\textcolor{red}{$\cdots$}}
\put(-60,119){\text{\tiny \textsc{argo} \textsc{Poisson}}}
\put(-70,112){\textcolor{OliveGreen}{---}}
\put(-60,112){\text{\tiny \textsc{$N$body} \textsc{Haloes}}}
\put(-69.5,105){\textcolor{magenta}{- -}}
\put(-60,105){\text{\tiny \textsc{argo} \textsc{Poisson-unity}}}

\put(-208,135){\text{\tiny$10^3$}}
\put(-208,93){\text{\tiny$10^2$}}
\put(-208,50){\text{\tiny$10^1$}}

\put(-206,34){\text{\tiny\tiny$1.6$}}
\put(-206,29){\text{\tiny\tiny$1.4$}}
\put(-206,24){\text{\tiny\tiny$1.2$}}
\put(-200,20){\text{\tiny\tiny$1$}}
\put(-206,16){\text{\tiny\tiny$0.8$}}
\put(-206,11){\text{\tiny\tiny$0.6$}}
\put(-206,6){\text{\tiny\tiny$0.4$}}

\put(-220,90){\rotatebox[]{90}{$P(k)\,[{h^{-1}\, \rm Mpc}]^3$}}
\put(-112,-15){$k\,[h\,{\rm Mpc}^{-1}]$}

\put(-225,20){\rotatebox[]{90}{$P(k)/$}}
\put(-216,20){\rotatebox[]{90}{$P^{N{\rm body}}(k)$}}

\put(-126,-5){\text{\tiny$10^{-1}$}}
\put(-5,-5){\text{\tiny$1$}}

\end{tabular}
\caption{{\it Upper panel}: Power spectra for the halo catalogue in green, the dark matter power spectrum in black, and the \textsc{argo} reconstructions:  mean Poisson reconstruction with unity bias in magenta (with negligible variance), mean Poisson reconstruction including a scale dependent bias in red (with 1$\,\sigma$ variance in filled yellow), and mean NB reconstruction including scale dependent bias in dashed  blue (with 1$\,\sigma$ standard deviation in filled cyan).  The mean power spectra and corresponding variances were estimated based on 10000 HMC iterations. {\it Lower panel}: Quotient between the reconstructed power spectra and the dark matter power spectrum using the same colour code as above. The power spectrum of the halo catalogue has been corrected for shot-noise and deconvolved with the NGP mass assignment kernel.}
\label{fig:power} 
\end{figure}

\begin{figure}
\begin{tabular}{c}
\hspace{0.5cm}
\includegraphics[width=7.cm]{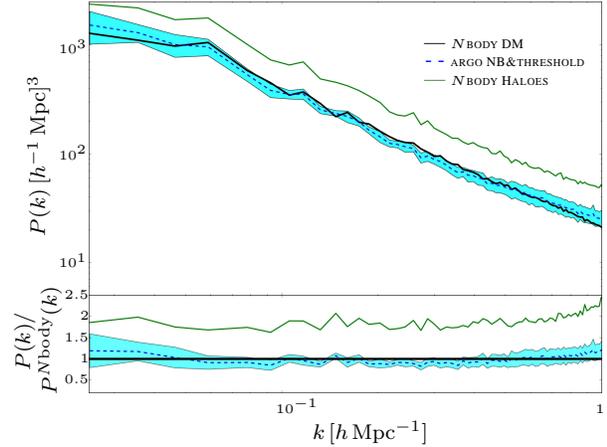}


\put(-70,133){\textcolor{black}{---}}
\put(-60,133){\text{\tiny \textsc{$N$body DM}}}
\put(-69.5,126){\textcolor{blue}{- -}}
\put(-60,126){\text{\tiny \textsc{argo} \textsc{NB\&threshold}}}
\put(-70,119){\textcolor{OliveGreen}{---}}
\put(-60,119){\text{\tiny \textsc{$N$body} \textsc{Haloes}}}

\put(-208,131){\text{\tiny$10^3$}}
\put(-208,91){\text{\tiny$10^2$}}
\put(-208,50){\text{\tiny$10^1$}}

\put(-206,38){\text{\tiny\tiny$2.5$}}
\put(-200,30){\text{\tiny\tiny$2$}}
\put(-206,22){\text{\tiny\tiny$1.5$}}
\put(-200,14){\text{\tiny\tiny$1$}}
\put(-206,6){\text{\tiny\tiny$0.5$}}

\put(-220,90){\rotatebox[]{90}{$P(k)\,[{h^{-1}\, \rm Mpc}]^3$}}
\put(-112,-15){$k\,[h\,{\rm Mpc}^{-1}]$}

\put(-225,20){\rotatebox[]{90}{$P(k)/$}}
\put(-216,20){\rotatebox[]{90}{$P^{N{\rm body}}(k)$}}

\put(-126,-5){\text{\tiny$10^{-1}$}}
\put(-5,-5){\text{\tiny$1$}}

\end{tabular}
\caption{{\it Upper panel}: Power spectra for the halo catalogue created by applying a mass cut in green, the dark matter power spectrum in black, and the \textsc{argo} reconstruction including scale dependent bias and thresholding of the matter overdensity in dashed  blue (with 1$\,\sigma$ standard deviation in filled cyan). The mean power spectra and corresponding variances were estimated based on 10000 HMC iterations. {\it Lower panel}: Quotient between the reconstructed power spectra and the dark matter power spectrum using the same colour code as above. The power spectrum of the halo catalogue has been corrected for shot-noise and deconvolved with the NGP mass assignment kernel.}
\label{fig:power2} 
\end{figure}

\section{Validation of the method}
\label{sec:results}

\begin{figure*}
\begin{tabular}{ccc}
\includegraphics[width=5.0cm]{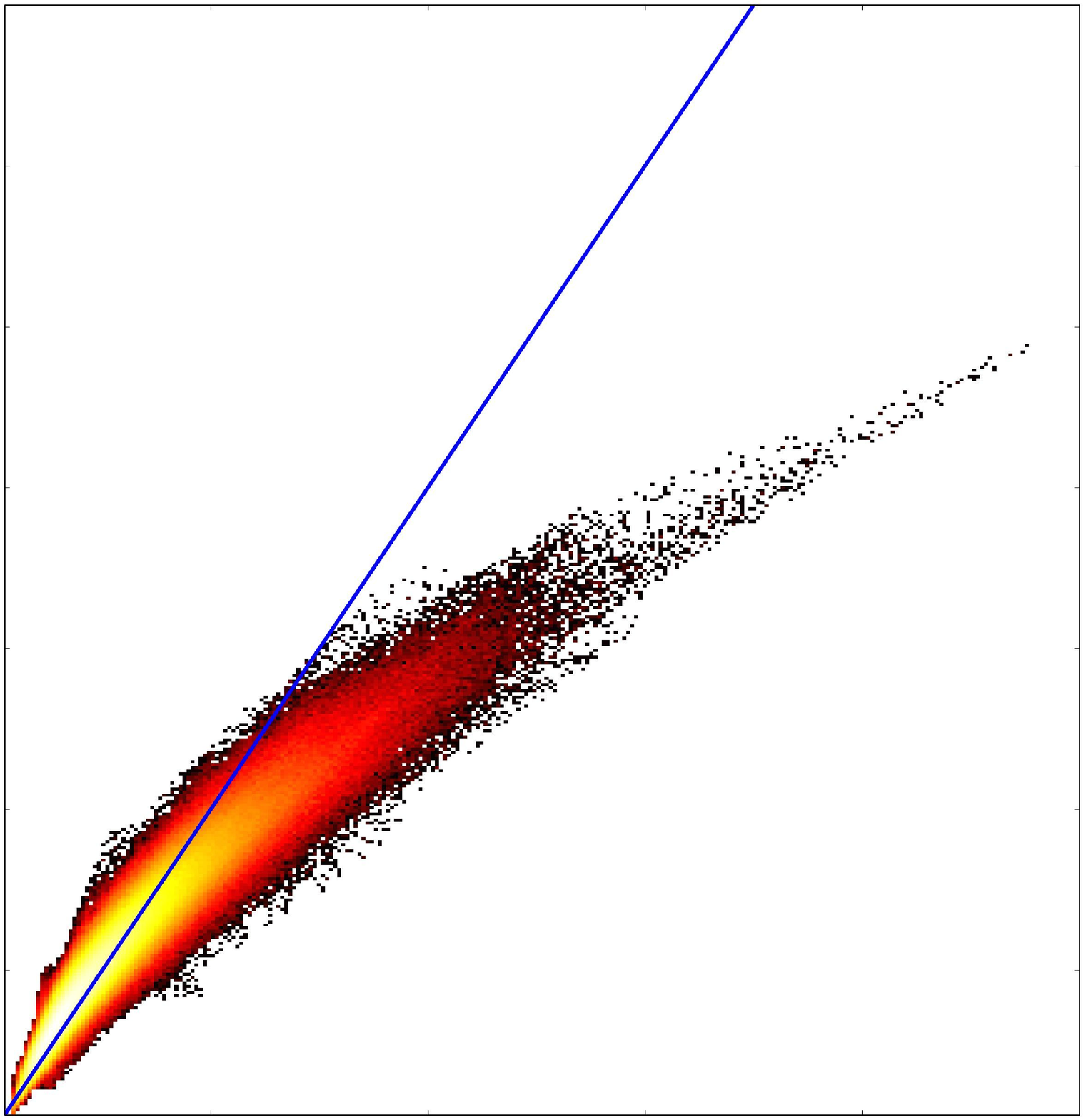}
\put(-75,20){\text{$ \delta_{\rm{x}}= \delta_{\textsc{Haloes}}^{\textsc{$N$body}}$}}
\put(-160,70){\rotatebox[]{90}{\text{{$1+\delta_{\rm{x}}$}}}}
\put(-148,142){\text{7}}
\put(-148,122){\text{6}}
\put(-148,102){\text{5}}
\put(-148,81){\text{4}}
\put(-148,59){\text{3}} 
\put(-148,38){\text{2}}
\put(-148,18){\text{1}}
\put(-148,-1){\text{0}}

\hspace{.1cm}
\includegraphics[width=5.0cm]{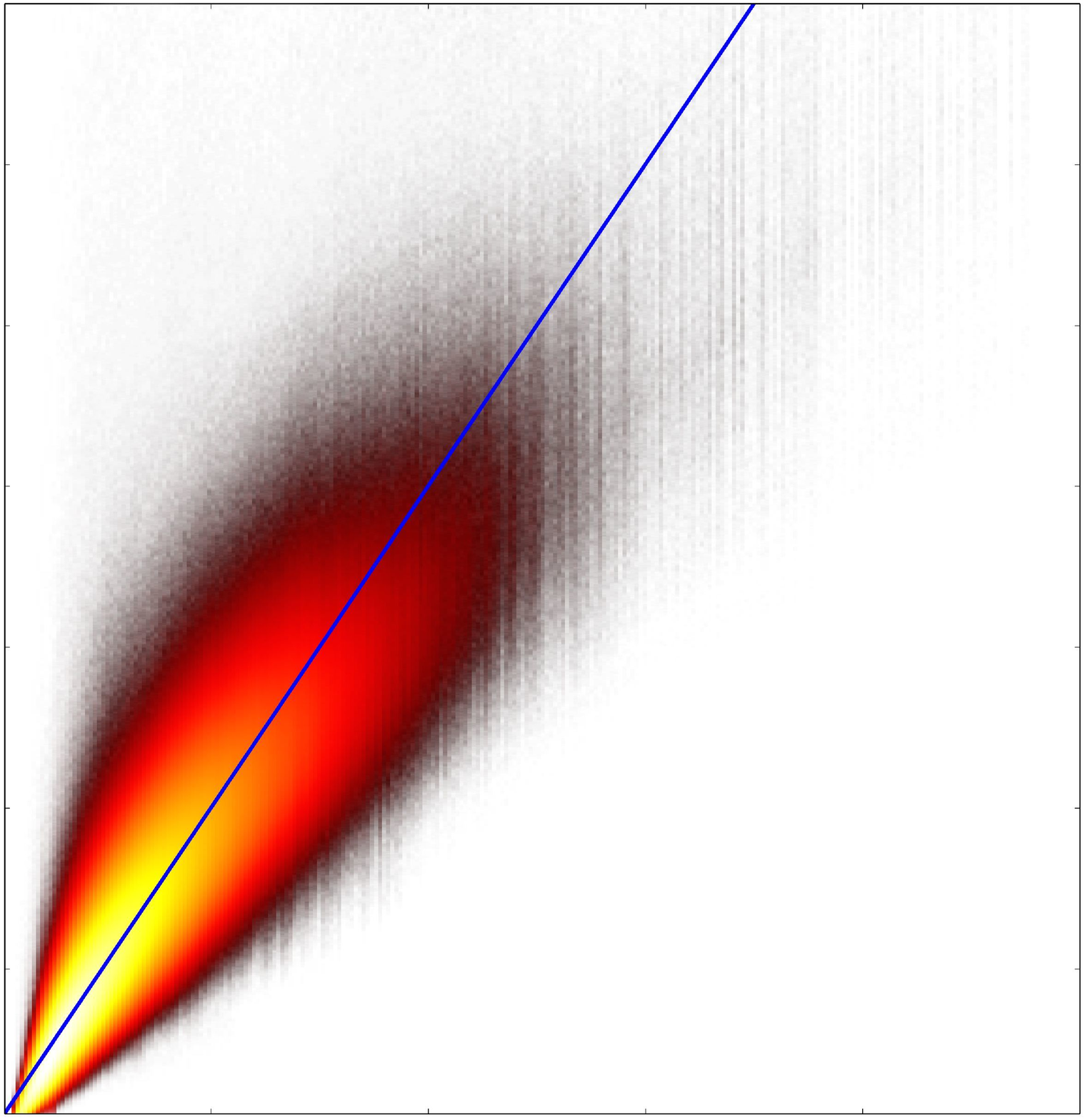}
\put(-75,20){\text{$ \delta_{\rm{x}}= \{\delta_{\textsc{NB}}^{\tiny\textsc{argo}}\}$}}

\includegraphics[width=0.388cm]{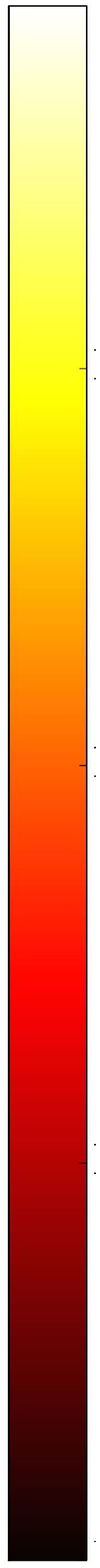}
\put(0,109){\text{$10^3$}}
\put(0,73){\text{$10^2$}}
\put(0,35){\text{10}}
\put(0,-1){\text{1}}
\hspace{.1cm}
\\
\includegraphics[width=5.1cm]{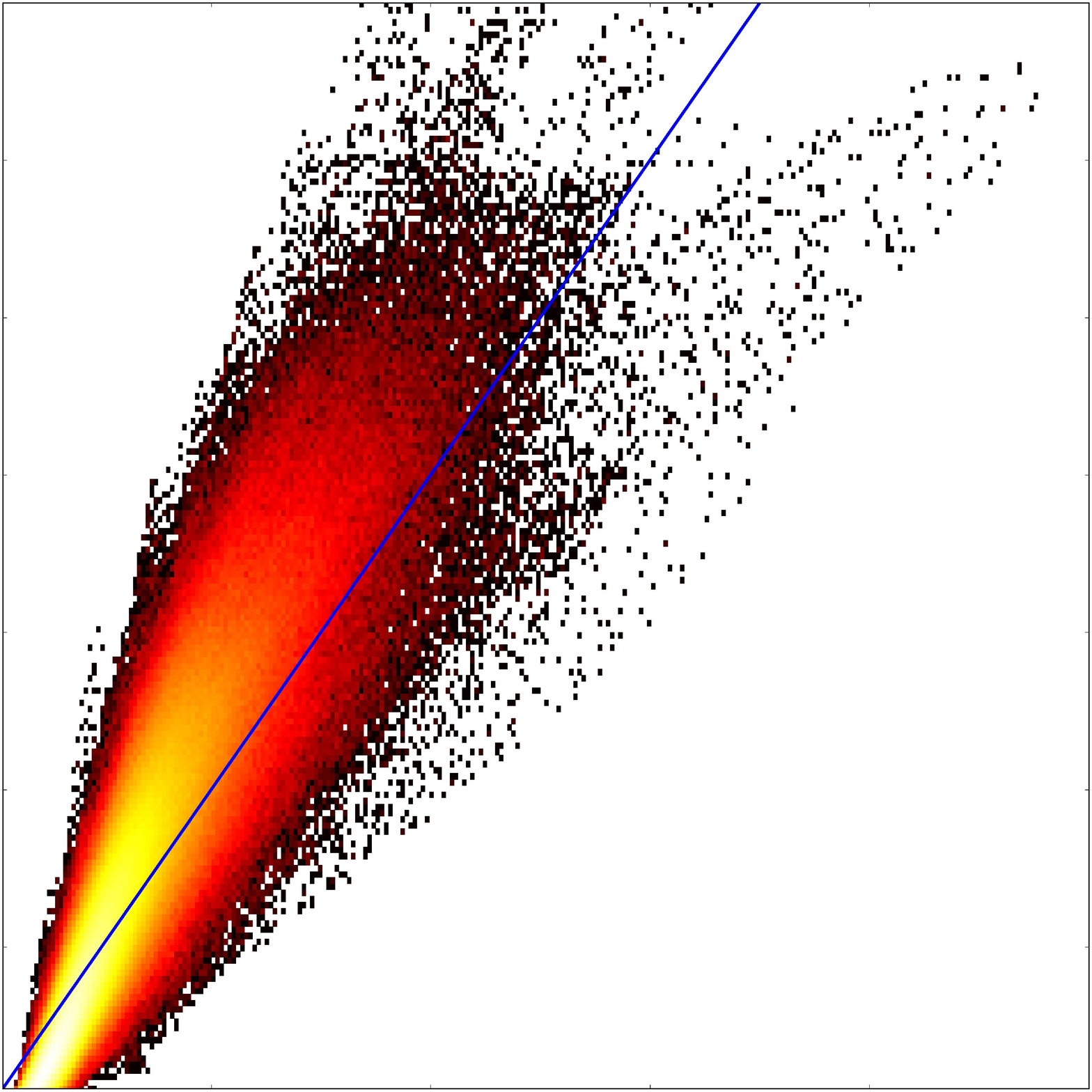}
\put(-160,70){\rotatebox[]{90}{\text{{$1+\delta_{\rm{x}}$}}}}

\put(-85,20){\text{$ \delta_{\rm{x}}= \{\delta_{\textsc{Haloes-Masscut}}^{\tiny\textsc{$N$body}}\}$}}
\put(-95,-17){\text{{ $1+\delta_{\textsc{M}}^{\textsc{$N$body}}$}}}
\put(-117,-7){\text{2}}
\put(-88,-7){\text{4}}
\put(-60,-7){\text{6}}
\put(-31,-7){\text{8}}
\put(-6,-7){\text{10}}
\hspace{0.cm}
\put(-148,142){\text{7}}
\put(-148,122){\text{6}}
\put(-148,102){\text{5}}
\put(-148,81){\text{4}}
\put(-148,59){\text{3}} 
\put(-148,38){\text{2}}
\put(-148,18){\text{1}}
\put(-148,-1){\text{0}}
\put(-117,-7){\text{2}}
\put(-88,-7){\text{4}}
\put(-60,-7){\text{6}}
\put(-31,-7){\text{8}}
\put(-6,-7){\text{10}}
\hspace{.1cm}
\includegraphics[width=5.1cm]{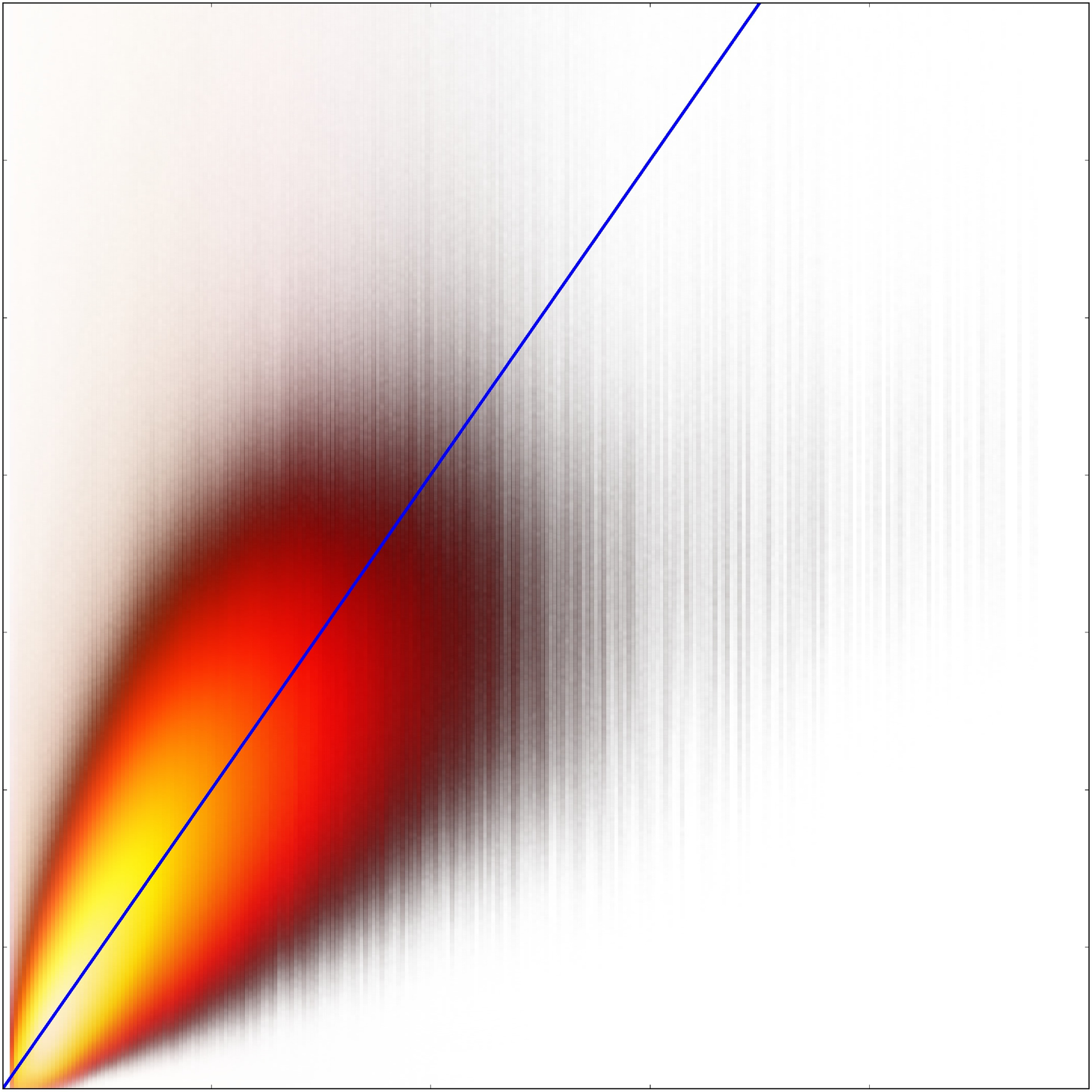}
\put(-75,20){\text{$ \delta_{\rm{x}}= \{\delta_{\textsc{NB-Masscut}}^{\tiny\textsc{argo}}\}$}}
\put(-95,-17){\text{{ $1+\delta_{\textsc{M}}^{\textsc{$N$body}}$}}}
\put(-117,-7){\text{2}}
\put(-88,-7){\text{4}}
\put(-60,-7){\text{6}}
\put(-31,-7){\text{8}}
\put(-6,-7){\text{10}}
\hspace{0.cm}
\includegraphics[width=0.388cm]{cbb}
\put(0,109){\text{$10^3$}}
\put(0,73){\text{$10^2$}}
\put(0,35){\text{10}}
\put(0,-1){\text{1}}
\end{tabular}

\caption{Cell-to-cell density correlation after Gaussian smoothing with radius $\sigma_{\rm{R}}=6$ {\hmpcnosp} between the dark matter density $1+\delta_{\textsc{M}}^{\textsc{$N$body}}$ and {\it top left panel}: the halo field with $2\times 10^5$ tracers, {\it top right panel}: the corresponding \textsc{argo} NB reconstruction, and {\it bottom left panel}: the halo field with $3\times 10^4$ tracers and mass cut, {\it bottom right panel}: the corresponding \textsc{argo} reconstruction with density thresholding. The colour code indicates the number of cells. In appendix \ref{app:haloes}, we also show the cell-to-cell density correlation after applying a unity bias Poisson reconstruction.}
\label{fig:c2c} 
\end{figure*}

To validate our algorithm we take the  Bolshoi dark matter simulation \citep{bolshoi} at redshift $z=0$, which was created using following cosmological parameters:  $\{ \Omega_{\Lambda}=0.73, \Omega_{\rm B}=0.047, \Omega_{\rm M}= 0.27,$ $\sigma_8 =  0.8, n_{\rm s}=0.95, H_0 = 100~ h~ \rm{km}~\rm{s}^{-1},$ $h=0.7 \} $.  

We consider two different subsamples of the halo catalogue, named $S_1$ and $S_2$, created with the bound-density-maxima \textsc{BDM} halo finder \cite{klypin1997} as inputs to our method.

For $S_1$ we randomly select $2\times 10^5$ haloes of the masses between $10^9$ to $10^{15}$ $\rm M_{\odot}$. These mass limits represent the total mass range of the Boshoi simulation (see appendix \ref{app:mass}) and yield a bias lower than one (see Fig. \ref{fig:power})

Additionally we consider the subsample $S_2$ with a mass cut of $3\cdot10^{12}\, \rm M_{\odot}$ (see appendix \ref{app:mass}), resulting in $3\times 10^4$ haloes and yielding a bias larger than one (see Fig. \ref{fig:power2}). 
Subsample $S_2$ permits us to test threshold bias described in Eq. \ref{eq:detbias}, which is negligible for subsample $S_1$. To technically overcome divergencies caused by the threshold bias we introduce a Gibbs sampling procedure, described in appendix \ref{app:gibbs} 

We use the nearest-grid-point (NGP) scheme to compute the number density for each cell on a mesh with $128^3$ cells. 
We use the  dark matter distribution from the Bolshoi simulation as a reference to estimate the accuracy of our reconstructions and define it as the {\it true} dark matter field.

To test the different models described in \S \ref{sec:bayes}, we apply our
novel \textsc{argo} code running a series of Hamiltonian Markov chains with
different likelihoods and bias assumptions: 
\begin{enumerate}
\item Poisson likelihood and unity bias using subsample $S_1$,
\item Poisson likelihood and power-law bias using subsample $S_1$,
\item NB likelihood and power-law bias using subsample $S_1$, 
\item NB likelihood and power-law bias including thresholding using subsample $S_2$. 
\end{enumerate}
We disregard the first 2000 iterations of the chains until the power spectra have converged and use a total of 10000 iterations for our analysis. The convergence behavior is estimated through the \cite{Gelman92} test in the appendix (see \S \ref{app:conv}).

Fig.~\ref{fig:slice} shows a slice through the volume of the Bolshoi simulation. This figure illustrates the problem, as the haloes (middle panel) represent in our case-study only 4 to 5 orders of magnitude less matter tracers than the dark matter particles used for the simulation (left panel). We have plotted the means of an ensemble of 10000 reconstructions using the NB model including a power-law bias (upper right panel without and lower right panel with thresholding bias) and find that the relevant structures are in general terms very well recovered. The filaments in the low density regions are less accurately reconstructed (see upper and lower right panel). This is expected from the low signal-to-noise ratio in those regions. Moreover, the models we are using for the low density regime are not optimal. We lack a good description of the dark matter field in the low density regime due to the lognormal approximation.  We also find that the high density peaks are less pronounced than in the {\it true} dark matter field. This is caused by the smoothing introduced from averaging the reconstructed samples, i.e., the mean estimator yields, as expected, a conservative reconstruction.
\subsection{Two-point statistics: power spectra}
\label{sec:two-point}
Let us further investigate the two-point statistics of the reconstructed fields in Fourier space, i.e., the power spectrum. This is shown in Fig.~\ref{fig:power} and Fig.~\ref{fig:power2}. 
In these plots we can clearly see the scale-dependent bias of the halo samples (solid green line) with respect to the dark matter field (solid black line). The power spectra of the halo fields have been corrected for the mass assignment kernel and shot-noise following \cite{2005ApJ...620..559J}. Therefore, the halo power spectra can be trusted up to approximatly $0.8\,h\,\rm{Mpc}^{-1}$, which is about 50\% of the Nyquist frequency. In Fig.~\ref{fig:power} the reconstruction using the Poisson likelihood with unity bias (dashed magenta line) is very close to the halo power spectrum, esentially confirming that the shot-noise correction follows a Poisson model, as the deterministic bias is neglected in this run.
We find a considerable improvement by adjusting the power-law bias (dotted red line). Nevertheless, the power spectrum  lackes power towards high $k$s, since the dispersion has been underestimated. We register $>$10\% deviation at $k=0.2\,h$ Mpc$^{-1}$.
Only when modeling also over-dispersion with the NB likelihood we find an excellent agreement with the dark matter power spectrum up to high $k$s of nearly 1$\,h\,$Mpc$^{-1}$. This is in agreement with the findings of \cite[][]{2014MNRAS.439L..21K}, however, by performing an inference analysis.
We also find that the power spectra show a larger scatter for the NB than for the Poisson case (compare cyan with yellow shaded regions, respectively). This is expected from the larger dispersion of the NB model.
\subsection{Cell-to-cell cross correlation}
\label{sec:c2c}
To further  assess the accuracy of our reconstructions, we compare them to the {\it true} dark matter field within a cell-to-cell correlation (see Fig.~\ref{fig:c2c}).  
To have a fair comparison we smooth each catalogue with a Gaussian kernel with smoothing length of $\sigma_{\rm{R}}=6$ \hmpcnosp  (see left panels in Fig. \ref{fig:c2c}). This should compensate for the different number density of haloes and dark matter particles. For the \textsc{argo} reconstructions the average over 10000 samples is shown.
We find that the haloes and the Poisson unity bias case show very similar cell-to-cell correlations, which are strongly biased towards high densities. This could have interesting applications to reconstruct the expected halo density field (see appendix \ref{app:haloes}, Fig. \ref{fig:unity-poisson})
In the low density regime we can see that the reconstruction is more biased than the halo sample. This must be caused by the lognormal assumption, as the bias was set to one in this run. The negative binomial reconstruction shows unbiased cell-to-cell correlations towards high densities, and the same bias at low densities. We can see that the scatter is larger for the NB than for the Poisson unity bias case or the halo distribution. This is expected from the  stochasticity included in the model. Non-local contributions, which may be deterministic, are absorbed in our stochastic bias model, thereby, possibly enhancing the scatter.
Interestingly, we see in our run (iv) including the thresholding bias, that the biased low halo density  (see lower left panel in Fig. \ref{fig:c2c}) is corrected in the low density region  (see lower right panel in Fig. \ref{fig:c2c}), achieving a similar accuracy to the run (iii) where haloes are included in the low density regions (see upper right panel in Fig. \ref{fig:c2c}). We also find in run (iv) using subsample $S_2$ (moderate massive objects of minimum mass of $3\cdot10^{12}\, \rm M_{\odot}$), that the deviation from Poissonity is negligible. We expect an under-dispersed distribution for more massive haloes, clusters or quasars. The method presented in this work is therefore optimal for over-dispersed tracers like emission line galaxies.

\section{Conclusions}
\label{sec:conc}

We have presented a Bayesian reconstruction algorithm, which is able to produce unbiased samples of the underlying dark matter field from non-linear stochastic biased tracers.

We find in this study that an accurate reconstruction of the dark matter field, requires both modeling of the stochastic and the deterministic bias. Our results using a power-law bias model and the negative binomial distribution function are very encouraging, as they produce unbiased statistics over a wide range of scales and density regimes.

We have focused on the negative binomial distribution function and discussed for which parameter dependencies it can be equivalent to the gravitational thermodynamics PDF.
Furthermore, we model the deterministic bias  relating the expected galaxy number density to the dark matter density through a non-linear scale-dependent expression.

We have presented the Hamiltonian equations of motions for our model and implemented them in the \textsc{argo} code. We  have shown that this permits us to efficiently sample the posterior distribution function of density fields given a set of biased tracers. 
We have also introduced a Gibbs-sampling scheme to deal with strongly biased objects tracing the high density peaks. 

In particular, we have tested our algorithm with the Bolshoi $N$-body simulation, inferring the underlying dark matter density field from a subsample of the corresponding halo catalogue. 
We found that a Poisson likelihood yields reconstructions with power spectra deviating more than 10\% at $k=0.2$ $h$ Mpc$^{-1}$. Our method shows that we can draw closely unbiased samples (compatible within 1 $\sigma$) from the posterior distribution up to scales of about $k\sim$1 $h$ Mpc$^{-1}$ in terms of power-spectra and cell-to-cell correlations  with the NB PDF. 

We have furthermore analytically shown that our method can deal with incomplete data and perform a multi-tracer analysis. Further investigation need to be done here to demonstrate the level of accuracy of such approaches.

We will demonstrate in a forthcoming publication that we can also correct for redshift space distortions in an iterative way.
Our work represents the first attempt to deal with nonlinear stochastic bias in a reconstruction algorithm and will contribute towards an optimal analysis of galaxy surveys.

\section*{Acknowledgments}
MA thanks the Friedrich-Ebert-Foundation.
The authors thank Benjamin Granett for encouraging discussions on the multi-tracer analysis.
The MultiDark Database used in this paper and the web application providing online access to it were constructed as part of the activities of the German Astrophysical Virtual Observatory as result of a collaboration between the Leibniz-Institute for Astrophysics Potsdam (AIP) and the Spanish MultiDark Consolider Project CSD2009-00064. 

{\small
\bibliographystyle{mn2e}
\bibliography{lit}
}

\appendix

\label{app}

\section{Multi-tracer analysis}
\label{sec:multi}

Let us assume that we have a set of $N$ halo or galaxy samples $\mbi N_{\rm G1},\dots,\mbi N_{{\rm G}N}$. 
We can write the joint problem of inferring the dark matter field conditioned on the different halo/galaxy samples by the following posterior PDF
\ba
\lefteqn{\mathcal P(\mbi\delta_{\rm M}|\mbi N_{{\rm G}1},\dots,\mbi N_{{\rm G}N},\{p_{\rm C}\},\mathcal B(\mbi N_{\rm G1},\dots,\mbi N_{{\rm G}N}|\mbi\delta_{\rm M}))\propto}\nonumber\\&&\pi(\mbi\delta_{\rm M}|\{p_{\rm C}\})\times\mathcal L(\mbi N_{\rm G1},\dots,\mbi N_{{\rm G}N}|\mbi \delta_{\rm M})\,,
\ea
with $\mathcal B(\mbi N_{\rm G1},\dots,\mbi N_{{\rm G}N}|\mbi\delta_{\rm M})$ being the joint bias.

If the samples have distinct biasing parameters, we can assume that each of the samples is conditioned on the underlying dark matter field only
\ba
\lefteqn{\mathcal L(\mbi N_{\rm G1},\dots,\mbi N_{{\rm G}N}|\mbi \delta_{\rm M},\mathcal B(\mbi N_{\rm G1},\dots,\mbi N_{{\rm G}N}|\mbi\delta_{\rm M}))\propto}\\&&\mathcal L(\mbi N_{\rm G1}|\mbi \delta_{\rm M},\mathcal B(\mbi N_{\rm G1}|\mbi\delta_{\rm M}))\dots\mathcal L(\mbi N_{{\rm G}N}|\mbi \delta_{\rm M},\mathcal B(\mbi N_{{\rm G}N}|\mbi\delta_{\rm M}))\nonumber\,.
\ea
Hence, we can write the posterior PDF as
\ba
\lefteqn{\mathcal P(\mbi\delta_{\rm M}|\mbi N_{\rm G1},\dots,\mbi N_{\rm GN},\mathcal B(\mbi N_{\rm G1}|\mbi\delta_{\rm M}),\dots,\mathcal B(\mbi N_{\rm GN}|\mbi\delta_{\rm M}))}\\&&\propto \pi(\mbi\delta_{\rm M}|\{p_{\rm C}\}) \nonumber\\&&\hspace{-0.5cm}\times \mathcal L(\mbi N_{\rm G1}|\mbi \delta_{\rm M},\mathcal B(\mbi N_{\rm G1}|\mbi\delta_{\rm M}))\dots\mathcal L(\mbi N_{\rm GN}|\mbi \delta_{\rm M},\mathcal B(\mbi N_{\rm GN}|\mbi\delta_{\rm M})) \nonumber\,.
\ea
For the Hamiltonian sampler (see \S~\ref{sec:ham}) we need to compute the potential energy, which is defined as
\ba
\lefteqn{-\ln\left(\mathcal P\left(\mbi\delta_{\rm M}|\mbi N_{\rm G1},\dots,\mbi N_{\rm GN},\mathcal B\left(\mbi N_{\rm G1}|\mbi\delta_{\rm M}\right),\dots,\mathcal B\left(\mbi N_{\rm GN}|\mbi\delta_{\rm M}\right)\right)\right)} \nonumber\\&&\hspace{-0.5cm} = {\rm const} -\ln\left(\pi\left(\mbi\delta_{\rm M}|\{p_{\rm C}\}\right)\right) \nonumber\\&&\hspace{-0.5cm}-\ln\left(\mathcal L\left(\mbi N_{\rm G1}|\mbi \delta_{\rm M},\mathcal B\left(\mbi N_{\rm G1}|\mbi\delta_{\rm M}\right)\right)\right) \nonumber\\&&\hspace{-0.5cm}\dots\nonumber\\&&\hspace{-0.5cm}-\ln\left(\mathcal L\left(\mbi N_{\rm GN}|\mbi \delta_{\rm M},\mathcal B\left(\mbi N_{\rm GN}|\mbi\delta_{\rm M}\right)\right)\right) \,.
\ea
This expression permits us to incorporate any additional galaxy sample and combine different galaxy catalogues with the method presented in this work.
The above calculations demonstrate that the dark matter field serves as a common denominator for different halo/galaxy samples and allows one to perform a multi-tracer analysis.

\section{Gradient calculations}
\label{app:grads}

Here we calculate the derivatives of the prior and the likelihood (see Section \ref{sec:sampling}) separately with signal vector $s_i$.
\begin{eqnarray}
- \frac{\partial}{\partial s_{i}}  \ln \mathcal{P}( \delta_{\mathrm{M}} | N_G) = &-& \frac{\partial}{\partial s_{i}} \ln  \pi(\delta_{\mathrm M} | \{p_C\})  \\
&-&  \frac{\partial}{\partial s_{i}} \ln \mathcal{L}( N_G | \lambda )
\end{eqnarray}
\subsection{Prior}
The derivative for the Gaussian prior for the linearized Gaussian field $\mbi s = \ln(1+\mbi\delta_{\rm M}^{\rm LN})-\mbi \mu$ writes as
\begin{eqnarray}
 - \frac{\partial}{\partial s_{i}} \ln \pi &=& - \frac{\partial}{\partial s_{i}}  \ln \left[ \frac{1}{\sqrt{(2\pi)^{N_{\mathrm{C}}}\mathrm{det}(\mathbf{S})}} \exp \left( -\frac{1}{2} \mbi s^{\dagger} \mathbf{S}^{-1} \mbi s\right)  \right] \nonumber  \\
 &=& \frac{1}{2}  \sum_{ij}(\delta_{ik} (S_{ij}^{-1}s_j)  +\delta_{jk} (s_i S_{ij}^{-1})  ) \nonumber \\
 &=& \frac{1}{2} \left[ \sum_j (S^{-1}_{kj} s_j) +\sum_i(s_i S^{-1}_{ik}  )\right ] \nonumber \\
  - \frac{\partial}{\partial s_{i}} \ln \pi &=&   \mathbf{S}^{-1} \mbi s
\end{eqnarray} 
\subsection{Likelihood}
We calculate the gradient of the likelihood
\be
\mathcal{L}_{\rm NB}=\prod_{i=1}^{N_C}\left[ \frac{\lambda_i^{N_i}}{N_i!} \frac{\Gamma(\beta+N_i)}{\Gamma(\beta)(\beta+\lambda_i)^{N_i}}\frac{1}{\left(1+\frac{\lambda_i}{\beta}\right)^\beta} \right]\,,
\ee
as follows
 \begin{flalign}
 \frac{\partial }{\partial s_i} &=  \left( \frac{\partial s_i}{\partial \delta_j}\right)^{-1} \frac{\partial \lambda_k}{\partial \delta_j} \frac{\partial }{\partial \lambda_k} \, ,   \\
 \left(  \frac{\partial s_i}{\partial \delta_j}\right)^{-1} &=1+\delta_j \, , \\
\frac{\partial \lambda_k}{\partial \delta_j} &= \frac{\alpha \lambda_j}{1+\delta_j}  \, .  
   \end{flalign}
   We can now just calculate the derivative of the likelihood w.r.t. $\lambda$ and get the final result as
   \begin{flalign}
  -\frac{\partial \ln \mathcal{L}_{\rm NB}}{\partial \lambda_k} &=-\frac{N_k}{\lambda_k}-\frac{N_k}{\beta + \lambda_k}+\frac{1}{1+\frac{\lambda_k}{\beta}} \,.
 \end{flalign}
Taking this into account, the derivative of the Likelihood writes as 
 \begin{flalign}
 -\frac{\partial  \ln \mathcal{L}_{\rm NB}}{\partial s_i} = \alpha \lambda_i \cdot \left[ -\frac{N}{\lambda_i}-\frac{N}{\beta + \lambda_i}+\frac{1}{1+\frac{\lambda_i}{\beta}}\right]
\end{flalign}
We note that for  $\lim_{\beta \to \infty} \left(-\frac{N}{\lambda}-\frac{N}{\beta + \lambda}+\frac{1}{1+\lambda/\beta} \right) = 1-\frac{N}{\lambda}$, which is identical to the solution of the 
likelihood.

\subsection{Gibbs sampling of the thresholding}
\label{app:gibbs}
The density cut-off bias component introduces a numerical instability, since the additional gradient terms diverge around the density threshold. Therefore we follow a Gibbs-sampling strategy by considering the $\theta(\delta_{\rm M}-\delta_{\rm th})$ step-function to be constant with respect to the signal $\delta_{\rm M}$. This permits us to neglect all the terms where gradients and logarithms of the step-function appear (see previous section). The scheme can be described as follows:
\begin{eqnarray}
 \delta_{\rm M} &\curvearrowleft& \mathcal{P}(\delta_{\rm M}| \theta,N_{\rm G}, \mathcal{M}(\delta_{\rm M}), \mathcal{B}(N_{\rm G}|\delta_{\rm M}))  \\
 \theta &\curvearrowleft&  \mathcal{P}(\theta|\delta_{\rm M},\delta_{\rm th}) 
\end{eqnarray}
We consider a vanishing uncertainty for the PDF of the step function given $\delta_{\rm M}$ and $\delta_{\rm th}$, which is equivalent to a Dirac delta density function. 
We have tested this scheme and found it very stable. Nevertheless, we consider investigating the introduction of some uncertainty in the PDF of the step function in future work. This may yield to an even better agreement in the power spectra.

\section{Convergence of the Hamiltonian sampler with nonlinear non-Poisson likelihoods}
\label{app:conv}
 
  \begin{figure}
  \begin{tabular}{c}
   \hspace{0.5cm}
   \includegraphics[width=7cm]{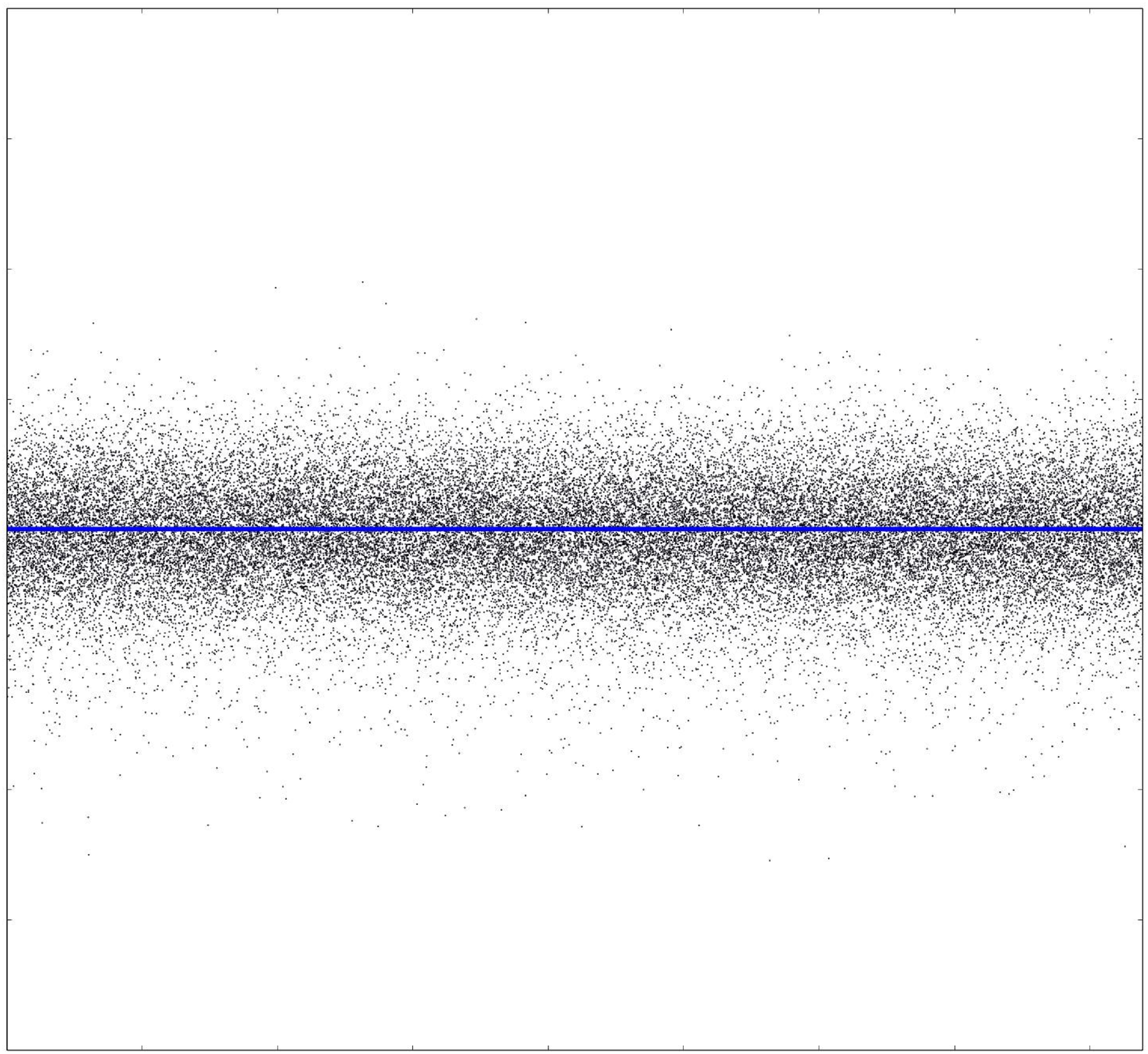}

\put(-210,175){\text{\tiny$1.20$}}
\put(-210,133){\text{\tiny$1.10$}}
\put(-210,89){\text{\tiny$1.00$}}
\put(-210,45.5){\text{\tiny$0.90$}}
\put(-210,4){\text{\tiny$0.80$}}

\put(-220,90){\rotatebox[]{90}{\it PSRF}}

\put(-158,-3){\text{\tiny$5\cdot 10^5$}}
\put(-113,-3){\text{\tiny$1\cdot 10^6$}}
\put(-64,-3){\text{\tiny$1.5\cdot 10^6$}}
\put(-19,-3){\text{\tiny$2\cdot 10^6$}}

\put(-110,-15){\text{\it Cell}}

\end{tabular}
 	\caption{{\it PSRF} from the Gelman-Rubin test for the converge of the Hamiltonian Markov Chain shown for the NB run including a power-law bias. Each point corresponds to one (out of $128^3$) specific cell of the reconstructed volume of \textsc{argo}. The {\it PSRF} estimator on the $y$-axis should be around $1$ to indicate that the chain is converged.}

 \label{fig:GR}
\end{figure}

There is no unique way to estimate the convergence of a Markov Chain nor a stopping criterion in literature. 
In principle, multiple chains are supposed to converge to the same stationary distribution and thus all samples should be consistent. Also within one running chain the drawn samples should be consistent after the burn-in phase. So comparing the means and variances within one converged chain to the samples of independently run chains will probe the convergence of the chains. This has been introduced  by \cite{Gelman92}. The test can be applied to any parameter without loss of generality.

We assume $N_{\rm chains}$ chains of length $N_{\rm length}$. The output of the chain is denoted as $x_{c,s}$, with $c\in \{1,2,...,N_{\rm chains}\}$ and $s\in \{1,2,...,N_{\rm length}\}$. In our case $x$ would the multidimensional ensemble $\{\delta_i\}$ of the overdensity of cell $i$ in our reconstructed volume. For simplicity we show the calculations for a one dimensional observable $x$. Starting from an identical proposal distribution we calculate as follows
\begin{enumerate}
 \item Calculate each chain's mean value
 \begin{align}
 \hspace{1cm} \bar x_c = \frac{1}{N_{\rm length}} \sum_{s} x_{c,s}\, . \nonumber
 \end{align}
 \item Calculate each chain's variance
 \begin{equation}
 \hspace{1cm} \sigma^2_c = \frac{1}{N_{\rm chains}-1} \sum_s (x_{c,s}-\bar x_c)^2 \, . \nonumber
 \end{equation}
  \item Calculate all chains' mean
 \begin{equation}
 \hspace{1cm} \bar x =  \frac{1}{N_{\rm chains}} \sum_{c} \frac{1}{N_{\rm length}} \sum_{s} x_{c,s} =  \frac{1}{N_{\rm chains}} \sum_{c} \bar x_c \, .\nonumber
 \end{equation}
  \item Calculate the weighted mean of each chain's variance
  \begin{equation}
 \hspace{1cm} B= \frac{N_{\rm length}}{N_{\rm chains}-1} \sum_{c} (\bar x_c -\bar x)^2 \, . \nonumber
 \end{equation}
   \item Calculate the average variance within one chain 
  \begin{equation}
 \hspace{1cm} W= \frac{1}{N_{\rm chains}} \sum_{c} \sigma^2_c \, . \nonumber
 \end{equation}
    \item The potential scale reduction factor ({\it PSRF}) then is defined as
  \begin{equation}
 \hspace{1cm}  {\rm \it PSRF} = \sqrt{\frac{N_{\rm length}-1}{N_{\rm length}} + \frac{N_{\rm chains}+1}{N_{\rm length}N_{\rm chains}}\frac{B}{W} }       \, . \nonumber
 \end{equation}
\end{enumerate}
If all chains converge to the same target distribution, we expect the variance within one chain to be close to the variance between the $N_{\rm chains}$ chains, so that the {\it PSRF} to be close to one.
\section{Mass ranges of the Bolshoi simulation}
\label{app:mass}
\begin{figure}
  \begin{tabular}{c}
   \hspace{0.5cm}
   \includegraphics[width=7cm]{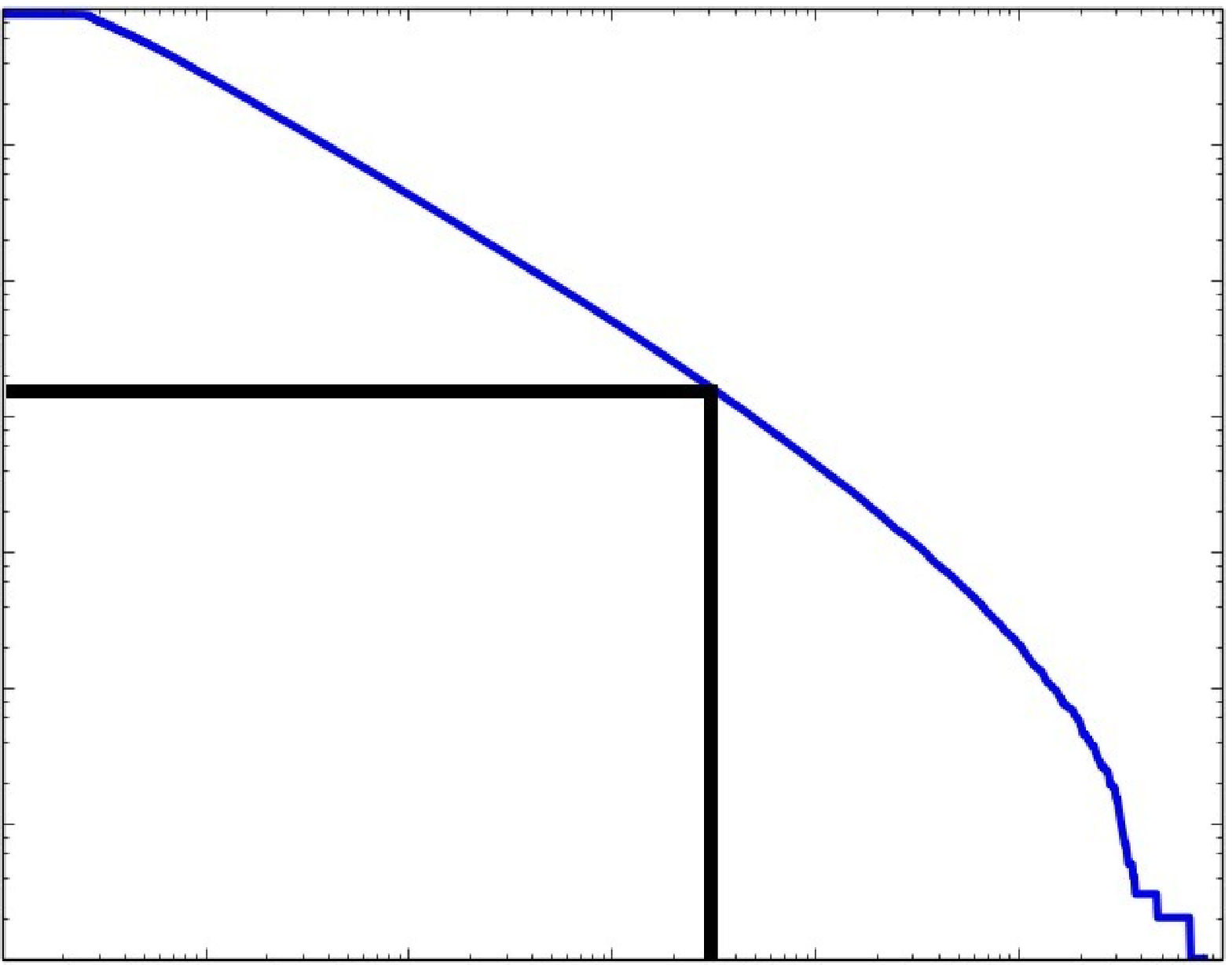}

\put(-210,129){\text{\tiny$10^6$}}
\put(-210,86){\text{\tiny$10^4$}}
\put(-210,43){\text{\tiny$10^2$}}
\put(-210,-1){\text{\tiny$10^0$}}

\put(-220,90){\rotatebox[]{90}{\it Number ($M>M_{\rm cut}$) }}

\put(-202,-6){\text{\tiny$10^9$}}
\put(-140,-6){\text{\tiny$10^{11}$}}
\put(-73,-6){\text{\tiny$10^{13}$}}
\put(-8,-6){\text{\tiny$10^{15}$}}

\put(-110,-17){\text{$M_{\rm cut}\,[M_{\odot}]$}}

\end{tabular}
 	\caption{Cumulative halo number densities for different lower halo mass cuts $M_{\rm cut}$.}

 \label{fig:mass_cut}
\end{figure}

In Fig. \ref{fig:mass_cut} we show the mass function of the Bolshoi simulation in a cumulative histogram. Our cut of $3\times 10^2 \rm M_{\odot}$ creates a subsample with 13000 tracers. 

\section{Reconstruction of the halo density field}
\label{app:haloes}
\begin{figure}
  \begin{tabular}{c}
   \hspace{0.5cm}
\includegraphics[width=5.0cm]{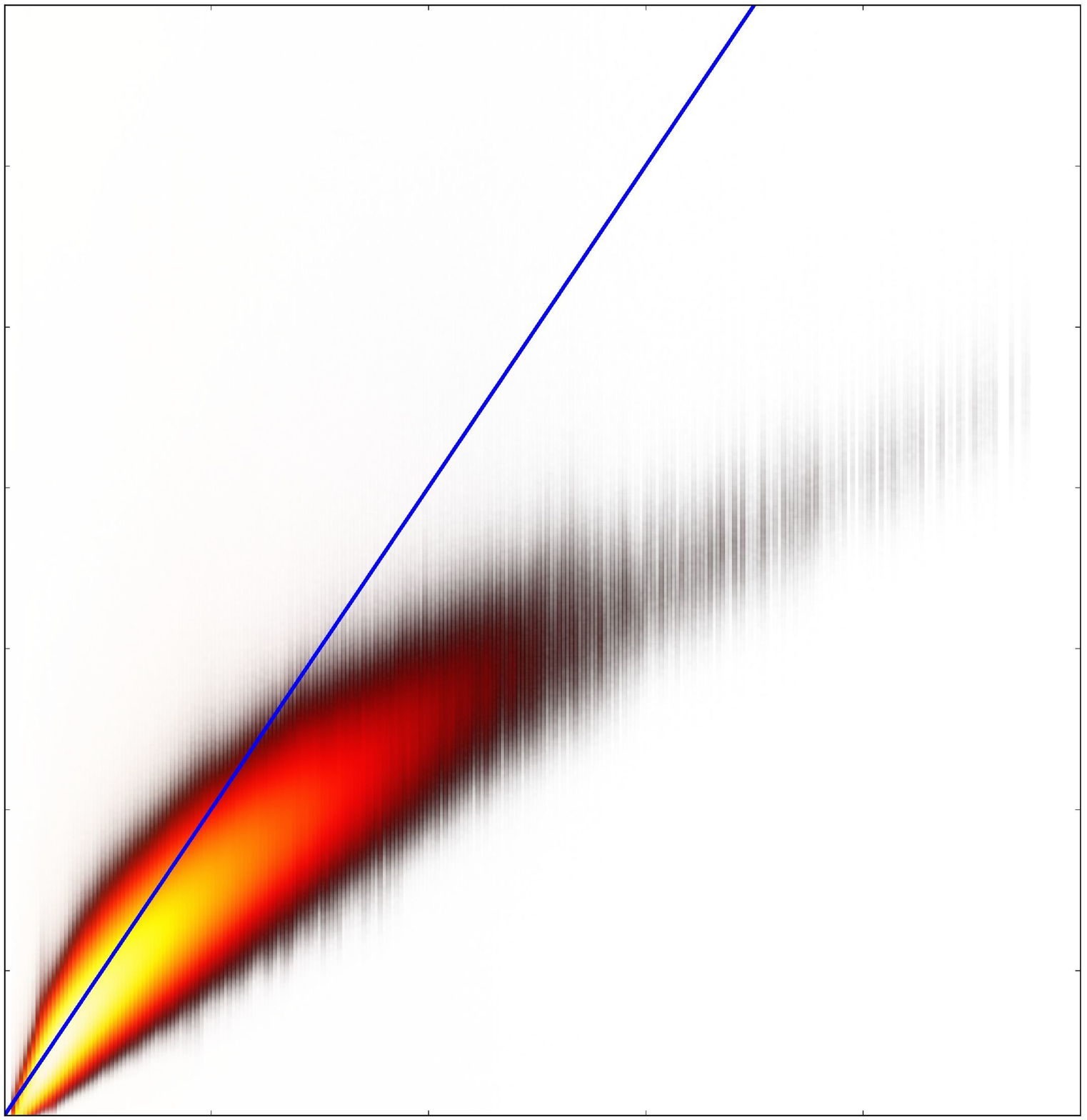}
\put(-75,20){\text{$ \delta_{\rm{x}}= \{\delta_{\textsc{Poisson-unity}}^{\tiny\textsc{argo}}\}$}}
\put(-95,-17){\text{{ $1+\delta_{\textsc{M}}^{\textsc{$N$body}}$}}}
\put(-160,70){\rotatebox[]{90}{\text{{$1+\delta_{\rm{x}}$}}}}
\put(-117,-7){\text{2}}
\put(-88,-7){\text{4}}
\put(-60,-7){\text{6}}
\put(-31,-7){\text{8}}
\put(-6,-7){\text{10}}

\includegraphics[width=0.388cm]{cbb}
\put(0,109){\text{$10^3$}}
\put(0,73){\text{$10^2$}}
\put(0,35){\text{10}}
\put(0,-1){\text{1}}
\end{tabular}
\caption{Cell-to-cell correlation shown a \textsc{argo} run with unity bias Poisson averaged over 10000 iterations.}

\label{fig:unity-poisson}
\end{figure}

We also run \textsc{argo} with a Poisson likelihood and unity bias. We can see in Fig. \ref{fig:unity-poisson} that this model creates smoothed reconstructions of the biased halo field (compare Fig. \ref{fig:c2c}). The averaged cell-to-cell correlation is in excellent agreement with the outcome of the halo field.

\end{document}